\documentclass[12pt]{article}
\usepackage{amssymb,amsmath,epsfig}
\usepackage{graphicx}
\usepackage{amsfonts,graphicx,amssymb,amsmath,amsthm,epsfig}
\usepackage{float}
\allowdisplaybreaks

\begin{document}
\title{\textbf{Cosmological Implications and Stability of $f\mathbb{(Q,T)}$ Gravity with
Pilgrim Dark Energy Model}}
\author{M. Sharif$^{1,2}$ \thanks{msharif.math@pu.edu.pk}~ and
\ Iqra Ibrar$^1$ \thanks {iqraibrar26@gmail.com}\\
$^1$ Department of Mathematics and Statistics, The University of Lahore,\\
1-KM Defence Road Lahore-54000, Pakistan.\\
$^2$ Research Center of Astrophysics and Cosmology, Khazar University,\\
Baku, AZ1096, 41 Mehseti Street, Azerbaijan.}

\date{}
\maketitle

\begin{abstract}
This manuscript endeavors to construct a pilgrim dark energy
framework within the $f\mathbb{(Q,T)}$ gravity theory, employing a
correspondence approach aligned with a non-interacting model that
incorporates pressureless matter alongside a power-law scale factor.
Here $\mathbb{Q}$ and $\mathbb{T}$ represent the non-metricity and
trace of the energy-momentum tensor, respectively. This extended
modified gravity framework accurately replicates various epochs in
the cosmological history. The $f\mathbb{(Q,T)}$ gravity models are
utilized to derive the equation of state parameter, phase planes and
squared speed of sound. The analysis reveals that the reconstructed
model exhibits an increasing or decreasing trend with the pilgrim
dark energy parameter. The equation of state parameter characterizes
the phantom regime, while the squared speed of sound parameter
provides a stable framework for examining the ongoing cosmic
evolution. The $\omega_{DE}-\omega'_{DE}$ plane trajectories reveal
the freezing region, while the $r-s$ phase plane shows the Chaplygin
gas model. It is important to highlight that our findings align with
the most recent observational data.
\end{abstract}
\textbf{Keywords}: Cosmic diagnostic parameters; Pilgrim dark energy
model; $f\mathbb{(Q,T)}$ gravity.\\
\textbf{PACS}: 64.30.+t; 95.36.+x; 04.50.kd.

\section{Introduction}

The general theory of relativity (GR) is a foundational pillar of
physics, revolutionizing our comprehension of gravity and the fabric
of spacetime. It is validated through comprehensive studies and
empirical evidences. However, this theory depends on the geometric
patterns found within Riemannian metric space. Weyl \cite{1-a}
devised an expansive geometric framework that surpasses Riemannian
space, offering a complete description of gravitational field and
matter. He focused on unifying gravitational and electromagnetic
forces, rather than attempting to combine all fundamental forces.
The Levi-Civita (LC) connection serves as a cornerstone in
Riemannian geometry, enabling the comparison of vectors based on
their lengths. In contrast, Weyl introduced a groundbreaking concept
that disregards vector magnitudes during parallel transport. To
address the absence of length information, he introduced an
additional component known as the length connection. Unlike the LC
connection, which preserves vector direction, the length connection
adjusts the conformal factor without affecting the direction of
transport. Non-Riemannian (NR) geometries extend the principles of
Riemannian geometry to provide a more nuanced understanding of
spacetime curvature. These geometrical frameworks incorporate
elements such as torsion and the non-metricity tensor
$(\mathbb{Q})$. According to Weyl's theory, the existence of
$\mathbb{Q}$ arises from the metric tensor possessing a non-zero
covariant derivative. \cite{1-b}.

The concept of $\mathbb{Q}$ within NR geometries offers an
alternative cosmological framework that eliminates the need for dark
energy (DE). In NR gravity models, the metric, coframe and full
connection serve as gauge potentials \cite{2-a}. The associated
field strengths include non-metricity tensor
$\mathbb{Q}_{\alpha\beta}$, torsion scalar $T$, and curvature tensor
$\mathbb{R}_{\alpha\beta}$. The NR gravity models are primarily
investigated due to the absence of experimental findings regarding
$\mathbb{Q}$ and torsion. The classification of spacetimes and
associated concepts is provided in Table $\textbf{1}$.

In STG, gravitational interactions are governed by the non-metricity
scalar, as opposed to the Ricci scalar $\mathbb{R}$ in general
relativity or the torsion scalar $T$ in teleparallel gravity. A
notable strength of using $\mathbb{Q}$ lies in its second-order
field equations, which avoid the higher-order derivatives common in
curvature-based models like $f\mathbb{(R,T)}$. This leads to
simpler, more stable cosmological formulations while still allowing
for modified dynamics. Additionally, $f\mathbb{(Q,T)}$ models can
reproduce cosmological observations such as large-scale structure
and redshift space distortions without relying on DE, presenting a
viable alternative to the $\Lambda$CDM paradigm. Non-metricity also
offers a cleaner geometric interpretation by separating
gravitational effects from curvature and torsion. In STG and its
extensions, spacetime is flat and torsionless, with gravity emerging
from variations in length and angle under parallel transport. This
provides a fully covariant and frame-independent formulation unlike
$f(T)$ theories, which suffer from frame dependence due to their
reliance on the tetrad formalism. Moreover, introducing the trace of
the EMT in $f\mathbb{(Q,T)}$ gravity enables non-minimal coupling
between matter and geometry. This coupling introduces new degrees of
freedom, offering greater flexibility for modeling late time cosmic
acceleration and the dynamics of DE without exotic fields. From a
cosmological standpoint, recent studies \cite{1x1} have shown that
$f\mathbb{(Q,T)}$ theories can successfully reproduce accelerated
expansion, phantom crossings, and even inflation, often
outperforming $f(\mathbb{R})$ and $f(T)$ models in observational
compatibility.

Scholars are increasingly attracted towards the study of NR
geometry, especially $f\mathbb{(Q)}$ theory, due to its potential to
reshape theoretical frameworks, its agreement with observational
evidence and its crucial role in advancing cosmological theories
\cite{2-b}. Barros et al. \cite{2-f} examined the cosmic properties
by utilizing redshift space distortion data within the framework of
non-metric gravity theory. The modified theory \cite{2-g} can
clarify the cosmic bounce scenario. Sharif and Ajmal \cite{2-gg}
examined the cosmological features of GGDE $f\mathbb{(Q)}$ gravity,
while Sharif et al. \cite{2-ggg} investigated the phenomenon of the
cyclic universe within NR geometry.
\begin{table}
\caption{Categorization of spacetimes}
\begin{center}
\begin{tabular}{|c|c|c|}
\hline Relations  & Spacetimes & physical representations
\\
\hline $\mathbb{Q}_{\mu\nu}=0,~ T=0,~ \mathbb{R}_{\mu\nu}=0$ &
Minkowski & Special Relativity
\\
$\mathbb{Q}_{\mu\nu}=0,~ T=0,~ \mathbb{R}_{\mu\nu}\neq0$ &
Riemannian & General Relativity
\\
$\mathbb{Q}_{\mu\nu}=0,~ T\neq0,~ \mathbb{R}_{\mu\nu}=0$ &
Weitzenb$\ddot{o}$ck & Teleparallel Gravity
\\
$\mathbb{Q}_{\mu\nu}\neq0,~ T=0,~ \mathbb{R}_{\mu\nu}=0$ & &
Symmetric Teleparallel
\\
$\mathbb{Q}_{\mu\nu}\neq0,~ T=0,~ \mathbb{R}_{\mu\nu}\neq0$ &
Riemann-Weyl & Einstein-Weyl
\\
$\mathbb{Q}_{\mu\nu}=0,~ T\neq0,~ \mathbb{R}_{\mu\nu}\neq0$ &
Riemann-Cartan & Einstein-Cartan
\\
$\mathbb{Q}_{\mu\nu}\neq0,~ T\neq0,~ \mathbb{R}_{\mu\nu}\neq0$ &
Non-Riemannian & Einstein-Cartan-Weyl
\\
\hline
\end{tabular}
\end{center}
\end{table}

Modified STG incorporates the trace of the EMT into the action
integral, resulting in the $f\mathbb{(Q,T)}$ theory. The inclusion
of $\mathbb{T}$ allows the gravitational field equations to directly
couple to the matter content of the universe, introducing additional
degrees of freedom. This coupling enables the model to reflect how
matter distribution influences spacetime geometry, potentially
explaining phenomena such as cosmic acceleration without requiring
exotic DE. Specifically, $\mathbb{T}$ encapsulates matter density
and pressure, providing a mechanism by which the universe's energy
distribution affects gravitational interactions. This approach
expands the scope of potential cosmological solutions while offering
a unified framework to address both early time inflation and the
late time acceleration of the cosmos. Unlike many modified gravity
theories that rely on extra fields or fine tuning, $f\mathbb{(Q,T)}$
extends GR naturally by modifying the spacetime geometry itself,
offering flexibility in fitting observational data and generating
new predictions for large scale cosmic behavior. With its
second-order field equations, this theory also simplifies
calculations, enhancing its practicality for theoretical studies.
While promising, $f\mathbb{(Q,T)}$ gravity is still emerging and
further studies are essential to fully understand its implications.

Recent research has begun to explore its cosmological applications,
demonstrating its potential to address fundamental questions about
the nature of DE and the universe accelerated expansion. Pati et al.
\cite{3-c} analyzed the evolutionary pattern of cosmological
parameters, observing a marginal change in the behavior of the
equation of state parameter (EoS). Narawade et al. \cite{3-f}
studied the dynamic behavior of an accelerating cosmological
paradigm in this gravity to confirm its stability. Venkatesha et al.
\cite{3-e} explored the existence of traversable wormhole solutions
within this extended theoretical framework. Gadbail et al. \cite{1a}
examined the evolution of the universe within this modified theory
to gain insights into accelerated expansion and other phenomena.
Bourakadi et al. \cite{1c} examined the impact of $f(\mathbb{Q},
\mathbb{T})$ gravity on the formation of primordial black holes.
Their research suggests that this theoretical framework could
significantly influence the dynamics of inflation. Shekh \cite{1d}
focused on late-time cosmic acceleration within the same framework,
proposed a new scale factor and conducted a statefinder analysis.
Gul et al. \cite{25} explored the feasibility and structural
integrity of compact stellar objects by considering various factors
within this theory. Khurana et al. \cite{26} explored a higher order
time-dependent function of the deceleration parameter with
restrictions from observations in the same theory.

Pati et al. \cite{A} explored little rip, big rip, and pseudo rip
cosmologies within extended STG using the form $f\mathbb{(Q,T)} =
a\mathbb{Q}^m + b\mathbb{T}$, deriving field equations, energy
conditions and cosmographic parameters in terms of non-metricity.
Agrawal et al. \cite{B} present a cosmological model based on a
generalized gravitational action, constraining its parameters with
current Hubble, Pantheon+ and Baryon Acoustic Oscillations data,
which reveals a universe evolving from early deceleration to
accelerating expansion and favoring a phantom field dominated phase.
Narawade et al. \cite{C} proposed an accelerating cosmological model
within extended STG, employing a hybrid scale factor constrained by
observational data, which reveals a transition from early-time
deceleration to late-time acceleration with quintessence-like
behavior approaching $\Lambda$CDM, while violating the strong energy
condition. Narawade et al. \cite{D} proposed an accelerating
cosmological model within extended STG, employing a hybrid scale
factor constrained by observational data, which reveals a transition
from early-time deceleration to late-time acceleration with
quintessence-like behavior approaching $\Lambda$CDM, while violating
the strong energy condition. Lohakare and Mishra \cite{E} proposed a
viable alternative to $\Lambda$CDM by analyzing $f(\mathbb{Q}, B)$
gravity through a dynamical system and Bayesian inference, revealing
a stable de Sitter phase and a smooth transition from deceleration
to late-time acceleration consistent with observational data.

The remarkable advancements in cosmology are evidenced by the
current rapid expansion of the cosmos. The change in the universe's
timeline, known as accelerated expansion, is driven by a paradigm
termed as an exotic force, which accommodates a repulsive nature
with significant negative pressure. This phenomenon is commonly
referred to as DE. It is assumed that this exotic energy will
determine the ultimate fate of the cosmos, yet its perplexing
characteristics remain unknown. The cosmic accelerating issue is
primarily addressed through two main approaches. Modified theories
of gravity represent one approach, serving as alternative frameworks
for incorporating DE.

Dynamical DE, a captivating idea in cosmology, has spurred
exploration of models to understand cosmic expansion. The alteration
of matter results not only in scalar field models like
\texttt{quintessence, k-essence and phantom fields} \cite{5-a}, but
also generates the dynamic DE models such as holographic, agegraphic
and pilgrim DE (PDE) models \cite{5-b}. The DE models at the
theoretical level remain consistent with observational data and can
be described using various energy density formulations. As a
consequence of energy dissipation, these models avoid the
coincidence issue that arises due to the dominance of DE. To prevent
the creation of black holes, Wei \cite{5-c} introduced the PDE
model, employing an energy density given by
\begin{equation}\label{A}
\rho_{DE}=3\alpha^{2}M_{p}^{4-u}H^{u},
\end{equation}
where $\alpha$ is non-zero constant, $H$ represents Hubble
parameter, $M_{p}$ denotes the Planck mass, and $u$ is the pilgrim
parameter.

The choice of PDE is rooted in its distinct theoretical foundation
and relevance to phantom cosmology. Unlike conventional holographic
DE models, PDE is designed specifically to account for a universe
dominated by phantom energy $\omega < -1$, a regime that
observational data from Planck and supernova surveys have
increasingly supported as a viable scenario for late-time cosmic
acceleration. The central idea behind PDE, is that strong phantom
energy not only drives accelerated expansion but also has sufficient
repulsive force to prevent the formation of black holes. This leads
to a scenario where DE acts as a powerful agent against
gravitational collapse, setting PDE apart from other DE models.
Furthermore, PDE naturally leads to a freezing region in the
$\omega_{DE} - \omega'_{DE}$ plane, which aligns with observational
features of DE evolution. Its inclusion in our model allows us to
examine whether a geometric theory of gravity specifically
$f\mathbb{(Q,T)}$ gravity that incorporates both non-metricity and
matter coupling via $\mathbb{T}$ can reproduce similar dynamics
without exotic fields. By reconstructing the PDE scenario under this
modified gravity setup, we provide a deeper insight into how
extended gravitational interactions and DE interplay to govern the
universe's evolution. This integration of PDE with $f\mathbb{(Q,T)}$
gravity not only enhances the richness of the theoretical model but
also contributes to a more unified framework that can potentially
explain both geometric and matter induced acceleration mechanisms.
Therefore, the adoption of PDE is both a natural and compelling
choice to explore within the scope of extended cosmological models.
Sharif and Zubair \cite{6-a} examined the cosmic expansion of the
PDE model using infrared (IR) cutoffs such as the event horizon,
particle horizon and the universe's conformal age. Jawad \cite{6-c}
examined the cosmological dynamics of the PDE model within the
framework of quantum cosmology, employing the Hubble horizon as an
infrared cutoff for the interacting framework.

Modified theories of gravity often utilize the reconstruction
phenomenon as a key approach for developing a DE model that can
precisely predict the trajectory of cosmic evolution. In this
context, comparisons are made between the respective energy
densities of DE and altered gravity theories. As a result, a desired
general function of the underlying gravity theory is derived using
the correspondence method of energy densities. Chattopadhyay et al.
\cite{6-f} delved into the enigmatic nature of the PDE model in
$f(T,T_{\mathbb{G}})$ ($\mathbb{G}$ is the \emph{Gauss-Bonnet
invariant}) gravity, concluding that an assertive phantom-like
behavior for $u=-2$ (PDE parameter) is crucial to avoid black hole
formation. Sharif and Rani \cite{6-d} constructed a PDE model based
on $f(T)$ gravity, aiming to describe the interactions of black
holes with phantom energy throughout the universe. The behavior of
the PDE model in $f\mathbb{(G)}$ gravity was examined \cite{6-b},
revealing that the dynamical framework uncovers multiple DE
scenarios. Sharif and Shah \cite{6-e} investigated the dynamics of
non-interacting and interacting PDE within a non-flat
Friedmann-Robertson-Walker (FRW) model using Brans-Dicke theory.
Myrzakulov et al. \cite{6-g} examined the dynamic behavior of the
EoS parameter, $\omega_{D}-\omega'_{D}$ and the statefinder planes
using the PDE model in $f\mathbb{(Q)}$ gravity.

This study utilizes a correspondence method to reconstruct the PDE
$f\mathbb{(Q,T)}$ model, with a particular emphasis on a
non-interacting scenario. We investigate cosmic evolution by
examining the EoS parameter and phase planes. The paper is
structured as follows: Section \textbf{2} introduces the
$f\mathbb{(Q,T)}$ gravity framework and its associated field
equations. Section \textbf{3} details the reconstruction approach
employed to develop the PDE $f\mathbb{(Q,T)}$ model. In section
\textbf{4}, we delve into cosmic dynamics by analyzing the EoS and
phase planes and we assess the stability of the model through the
squared speed of sound ($\nu_{s}^{2}$). The paper concludes with a
discussion of our research findings.

\section{Geometrical Basis of $f(\mathbb{Q,T})$ Gravity}

\section{Reconstruction of PDE $f\mathbb{(Q,T)}$ Model}

We consider the uniformly homogeneous and isotropic FRW universe
model, described by
\begin{equation}\label{33}
ds^{2} = -dt^{2}+ \mathbf{a}^{2}(t)[d\emph{x}^{2}+
d\emph{y}^{2}+d\emph{z}^{2}],
\end{equation}
where $\mathbf{a}$ represents the \emph{scale factor} of the
universe. The isotropic matter setup, specified through the
four-velocity $u_{\mu}$, along with the  normal matter density
($\rho_{M}$) and pressure ($P_{M}$) is expressed as follows
\begin{equation}\label{33a}
\tilde{\mathbb{T}}_{\mu\nu}=P_{M}g_{\mu\nu}+(\rho_{M}+P_{M})u_{\mu}u_{\nu}.
\end{equation}
The modified Friedmann equations within the framework of
$f\mathbb{(Q, T)}$ theory can be formulated  as
\begin{align}\label{34}
\rho_{M}+\rho_{DE}&=\rho_{eff}=3H^2,
\\\label{34-a}
P_{M}+P_{DE}&=P_{eff}=2\dot{H}+3H^2,
\end{align}
where $H =\frac{\dot{\mathbf{a}}}{\mathbf{a}}$, with the dot
notation indicating differentiation with respect to the universal
time $t$. The expression for $\mathbb{Q}$ in relation to the Hubble
parameter is given by
\begin{align}\label{80}
\mathbb{Q}=-\frac{1}{4}\big[-\mathbb{Q}_{\rho\xi\eta}\mathbb{Q}^{\rho\xi\eta}
+2\mathbb{Q}_{\rho\xi\eta}\mathbb{Q}^{\xi\rho\eta}-2\tilde{\mathbb{Q}}^{\rho}
\mathbb{Q}_{\rho}+\mathbb{Q}^{\rho}\mathbb{Q}_{\rho}\big],
\end{align}
which can be simplified to
\begin{equation}\label{86}
\mathbb{Q}=6H^{2}.
\end{equation}
Furthermore, $\rho_{DE}$ and $P_{DE}$ denote the DE density and
pressure and are given by
\begin{align}\label{35}
\rho_{DE}&=\frac{1}{2}f\mathbb{(Q,T)}
-f_{\mathbb{T}}(\rho_{M}+P_{M})-\mathbb{Q}f_{\mathbb{Q}},\\\label{36}
P_{DE}&=-\frac{1}{2}f\mathbb{(Q,T)}+2f_{\mathbb{Q}}\dot{H}
+\mathbb{Q}f_\mathbb{{Q}}+2f_{\mathbb{QQ}}H.
\end{align}

For an ideal fluid, the non-conservation equation \eqref{32} assumes
the following expression
\begin{equation}\label{37}
3H(\rho_{M}+P_{M})+\dot{\rho}_{M}=\frac{1}{(f_{\mathbb{T}}-1)}
\bigg[2\nabla_{\beta}(P_{M}\mu^{\beta}f_{\mathbb{T}})
+f_{\mathbb{T}}\nabla_{\beta}\mu^{\beta}\mathbb{T}
+2\mu^{\beta}\mathbb{T}_{\alpha\beta}\nabla^{\alpha}
f_{\mathbb{T}}\mu^{\beta}\bigg].
\end{equation}
With reference to the \eqref{34}, it follows that
\begin{equation}\label{38}
\Omega_{DE}+\Omega_{M}=1,
\end{equation}
In this framework, $\Omega_{M} = \frac{\rho_{M}}{3H^2}$ and
$\Omega_{DE} = \frac{\rho_{DE}}{3H^2}$ depict the fractional energy
densities of normal matter and DE, respectively. Cosmological models
that incorporate dynamic DE, whose energy density is directly
related to the Hubble parameter, are essential for understanding the
observed rapid expansion of the universe. In this study, we focus on
the non-interacting case to establish the core properties and
stability of the PDE model in $f\mathbb{(Q,T)}$ gravity. Moving
forward, we will apply a correspondence method within an ideal fluid
framework to develop the PDE $f\mathbb{(Q,T)}$ gravity model for a
dust scenario where $P_{M}= 0$

\subsection{Non-interacting PDE $f\mathbb{(Q,T)}$ Model}

In this section, we delve into the standard formulation of the
$f\mathbb{(Q,T)}$ function as outlined in \cite{10-c}
\begin{equation}\label{40}
f\mathbb{(Q,T)}=f_{1}\mathbb{(Q)}+f_{2}\mathbb{(T)}.
\end{equation}
Under these conditions, the interaction between the matter
components and curvature is minimal. This specific version of the
generic function denotes an entirely gravitational interaction,
making it easy to handle. This provides a detailed explanation for
the universe ongoing expansion. Furthermore, the reconstruction
process confirms that the resulting models are consistent with
physical principles \cite{10-c}. By applying Eq.\eqref{40}, the
field equations \eqref{34} and \eqref{34-a} lead to the following
\begin{equation}\label{41}
\rho_{M}+\rho_{DE}=\rho_{eff}=3H^{2}, \quad
P_{DE}=P_{eff}=2\dot{H}+3H^{2},
\end{equation}
where
\begin{align}\label{42}
\rho_{DE}&=\frac{1}{2}f_{1}\mathbb{(Q)}-\mathbb{Q}
f_{1\mathbb{Q}}+\frac{1}{2}f_{2}\mathbb{(T)}+
f_{2\mathbb{T}}\rho_{M}, \\\label{43}
P_{DE}&=-\frac{1}{2}f_{1}\mathbb{(Q)}-\frac{1}{2}f_{2}\mathbb{(T)}
+2f_{1\mathbb{Q}}\dot{H}+2
f_{1\mathbb{QQ}}H+\mathbb{Q}f_{1\mathbb{Q}}.
\end{align}
The related conservation equation \eqref{37} simplifies to
\begin{align}
\dot{\rho}_{M}+3H\rho_{M}=\frac{1}{f_{2\mathbb{T}}-1}\big[2\mathbb{T}
f_{2\mathbb{TT}}+f_{2\mathbb{T}}\dot{\mathbb{T}}\big].
\end{align}
This equation reduces to the conventional continuity equation when
the right-hand side is annulled, indicating that
\begin{equation}\label{45} 3H\rho_{M}+{\dot\rho_{M}}=0\quad\Longrightarrow\quad
\mathbf{a}(t)^{-3}\rho_{0}=\rho_{M},
\end{equation}
with the constraint
\begin{equation}\label{46}
f_{2\mathbb{T}}+2\mathbb{T}f_{2\mathbb{TT}}= 0,
\end{equation}
whose approach yields
\begin{equation}\label{47}
f_{2}(\mathbb{T})=B_{1}\mathbb{T}^\frac{1}{2}+B_{2},
\end{equation}
where $\rho_{0}$, $B_{1}$ and $B_{2}$ represent the integration
constants. For $M_{p}^{4-u}=1$ in Eq.\eqref{A}, we obtain
\begin{equation}\label{A-1}
\rho_{DE}=3\alpha^{2}H^{u}.
\end{equation}

The addition of the PDE model within the modified $f\mathbb{(Q,T)}$
gravity framework complements the geometrical modification of the
Einstein-Hilbert action by providing a practical mechanism for
detailed cosmic diagnostics and aligning with observational data on
DE. While the modified Einstein-Hilbert action addresses cosmic
acceleration theoretically, the PDE model enables a focused
examination of DE dynamics, such as the EoS and stability. By
allowing specific empirically testable scenarios, the PDE model
enhances the robustness and observational consistency of
$f\mathbb{(Q,T)}$ gravity, particularly in describing late-time
acceleration and phantom behavior, ultimately creating a more
comprehensive framework for understanding DE. We employ
Eqs.\eqref{42} and \eqref{A-1} along with the constraint on
$f_{2}(\mathbb{T})$ provided in \eqref{47} to develop a
reconstruction framework using the correspondence method
\begin{equation}\label{48}
\frac{f_{1}(\mathbb{Q})}{2}-\mathbb{Q}
f_{1\mathbb{Q}}+B_{1}\mathbb{T}^\frac{1}{2}+\frac{1}{2}B_{2}=3\alpha^{2}H^{u},
\end{equation}
or
\begin{equation}\label{48}
f_{1}(\mathbb{Q})-2\mathbb{Q}
f_{1\mathbb{Q}}+2B_{1}\mathbb{T}^\frac{1}{2}+B_{2}=6\alpha^{2}H^{u},
\end{equation}
where $g_{1}=6\alpha^{2}$ is an arbitrary constant.

We apply the \emph{power-law} solution for the \emph{scale factor},
presented as
\begin{equation}\label{49}
\mathbf{a}(t)=\mathbf{a}_{0}t^m,   \quad m > 0.
\end{equation}
Here, $\mathbf{a}_{0}$ denotes the present scale factor. In this
study, we have chosen the power-law form for the scale factor, for
several well-justified reasons. Mathematically, this form greatly
simplifies the structure of the field equations within the modified
gravity framework, particularly in the context of $f\mathbb{(Q,T)}$
gravity, enabling direct analytical solutions that are otherwise
challenging to obtain. This analytical tractability facilitates a
clearer and more systematic investigation of cosmic dynamics.
Additionally, power-law cosmology is widely used in gravitational
models, as it effectively encapsulates various key evolutionary
phases of the universe including radiation-dominated,
matter-dominated, and accelerated expansion epochs. Its flexibility
allows a smooth interpolation between these phases, providing a
robust and interpretable structure for modeling the behavior of DE.
Moreover, the power-law form proves advantageous for qualitative
studies, such as examining the evolution of the EoS, conducting
phase space analysis, and testing stability conditions. While it
represents a specific class of solutions, it serves as a reasonable
approximation for late-time cosmology and yields insights that are
qualitatively consistent with broader dynamical models. It also
supports effective comparison with observational data, thus
enhancing the model's testability and reliability.

Using this relationship, $H$, its derivative and the $\mathbb{Q}$
can be expressed as functions of cosmic time $t$ as
\begin{equation*}
\mathbb{Q}=6 \frac{m^2}{t^2}, \quad H=\frac{m}{t},
\quad\dot{H}=-\frac{m}{t^{2}}.
\end{equation*}
Substituting Eq.\eqref{49} into \eqref{45}, it can be concluded that
\begin{equation}\label{43-a}
\rho_{M}=\rho_{0}(\mathbf{a}_{0}t^m)^{-3}.
\end{equation}
By applying Eq.\eqref{49} to \eqref{48}, the function
$f_{1}(\mathbb{Q})$ can be determined as
\begin{align}\nonumber
f_{1}(\mathbb{Q})&=\sqrt{\mathbb{Q}} \bigg[4 \bigg\{\frac{a
\sqrt{\mathbb{T}} 2^{\frac{3 m}{8}+2} 3^{\frac{3 m}{8}}
\bigg(\frac{\mathbb{Q}}{m^2}\bigg)^{-\frac{3 m}{8}}}{(3 m+4)
\sqrt{\mathbb{Q}}}+\frac{g_1 6^{-u} m^{-u}
\mathbb{Q}^{u+\frac{1}{2}}}{2 u+1}\bigg\}+\frac{2
b}{\sqrt{\mathbb{Q}}}\bigg]
\\\label{50} &+g_2 \sqrt{\mathbb{Q}},
\end{align}
where $g_{2}$ is the integration constant, $a$ and $b$ are arbitrary
constant. As a result, the reconstructed $f\mathbb{(Q,T)}$ framework
is derived by inserting Eqs.\eqref{47} and \eqref{50} into
\eqref{40}, yielding the following expression
\begin{align}\nonumber
f\mathbb{(Q,T)}&=\sqrt{Q} \bigg[4 \bigg\{\frac{a \sqrt{\mathbb{T}}
2^{\frac{3 m}{8}+2} 3^{\frac{3 m}{8}}
\bigg(\frac{Q}{m^2}\bigg)^{-\frac{3 m}{8}}}{(3 m+4)
\sqrt{Q}}+\frac{g_1 6^{-u} m^{-u} Q^{u+\frac{1}{2}}}{2
u+1}\bigg\}+\frac{2 b}{\sqrt{Q}}\bigg]
\\\label{51}
&+B_{1} \sqrt{\mathbb{T}}+B_{2}+g_2 \sqrt{Q}.
\end{align}
For the sake of simplicity, we have taken $\mathbb{T}=\rho_{0}$.

Next, we express this function with respect to the redshift
parameter $z$. The deceleration parameter $q$ is subsequently
defined by
\begin{equation}\label{67}
q=-\frac{\mathbf{a}\ddot{\mathbf{a}}}{\dot{\mathbf{a}}^{2}}.
\end{equation}
The progression of the cosmic \emph{scale factor}, with respect to
$q$, is outlined as
\begin{equation}\label{68}
\mathbf{a}(t)=t^{(1+q)^{-1}},
\end{equation}
where $q=-0.831^{+0.091}_{-0.091}$ \cite{8a}. In this context, we
assume $\mathbf{a}_0$ to be unity \cite{8b}. The frequent use of the
\emph{power-law} form in cosmological models is justified for
several reasons. Its mathematical simplicity facilitates
straightforward analytical calculations, making it easier to derive
and analyze equations describing the universe's evolution. The
\emph{power-law} model indicates that the universe is expanding when
the $q$ is greater than -1. Moreover, \emph{power-law} solutions
naturally arise from the equations in some modified gravity theories
and specific cosmological models. The essential characteristics of
the expansion stage and the current cosmological development are
outlined below
\begin{equation}\label{69}
H=t^{-1}(1+q)^{-1}, \quad H_{0}=(q+1)^{-1}t^{-1}_{0}.
\end{equation}
Within \emph{power-law} cosmology, the expansion dynamics of the
cosmos are governed by two key factors: the Hubble constant $H_{0}$
and $q$. By exploring the relationship between $\mathbf{a}$ and $z$,
we are able to
\begin{equation}\label{70}
H=H_{0}\Upsilon^{1+q},
\end{equation}
here $\Upsilon=z+1$. Substituting Eq.\eqref{70} into \eqref{86},
$\mathbb{Q}$ can be expressed as follows
\begin{align}\label{71}
\mathbb{Q}=6H_{0}^{2}\Upsilon^{2+2q}.
\end{align}
\begin{figure}\center
\epsfig{file=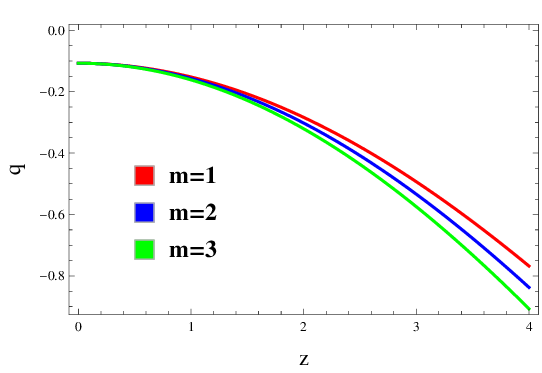,width=.5\linewidth}\caption{Graph of $q$ versus
$t$.}
\end{figure}

Figure \textbf{1} presents the deceleration parameter versus
redshift for three different values of $m$, showing a decreasing
trend with redshift. The parameter transitions into negative values,
which correspond to the accelerating phase of the universe. This
behavior aligns well with current observational data, thereby
supporting the physical validity of our model. Figure \textbf{2}
illustrates the evolution of the Hubble parameter with redshift for
different values of $m$. As expected, the Hubble parameter shows
positive behavior and an increasing trend with redshift, which is
consistent with observational data and confirms that the model
accurately captures the universe expansion dynamics. Substituting
this value into Eq.\eqref{51}, we obtain the reconstructed solution
for the PDE $f(\mathbb{Q,T})$ as it relates to the redshift
parameter, leading to
\begin{align}\nonumber
f\mathbb{(Q,T)}&=\frac{16 a \sqrt{\rho_{0}} \big(\frac{\beta ^2
\Upsilon^{2 q+2}}{m^2}\big)^{-\frac{3 m}{8}}}{3 m+4}+2 b+\frac{24
g_1 m^{-u} \big(\beta ^2 \Upsilon^{2 q+2}\big)^{u+1}}{2 u+1}
\\\label{52}
&+B_{1} \sqrt{\rho_{0}}+B_{2}+\sqrt{6} g_2 \sqrt{\beta ^2 \Upsilon^{2 q+2}}.
\end{align}
\begin{figure}\center
\epsfig{file=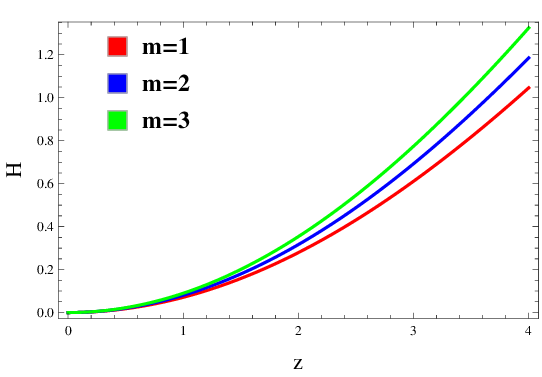,width=.5\linewidth}\caption{Graph of $H$ versus $t$.}
\end{figure}

In the graphical examination, the parameters are fixed at $a = 5$, $b = 1$ and $u =
\pm2$. As illustrated in Figure \textbf{3}, the reconstructed PDE $f\mathbb{(Q, T)}$
framework exhibits a consistent decline for $u = 2$ and an upward trend for $u = -2$.
Figures \textbf{4} and \textbf{5} illustrate the variation of $\rho_{DE}$ and
$P_{DE}$ with the redshift parameter. The $\rho_{DE}$ exhibits an increase, while
$P_{DE}$ stays negative across all values of $u$ and $m$, aligning with the expected
dynamics of DE. These attributes are further scrutinized within the reconstructed PDE
$f(\mathbb{Q,T})$ framework. By substituting \eqref{52} into \eqref{42} and
\eqref{43}, the following results are obtained
\begin{align}\nonumber \rho_{DE}&= 2 a \sqrt{\rho_{0}} \big(\frac{H_{0}^2
\Upsilon^{2 q+2}}{m^2}\big)^{-\frac{3 m}{8}}+b+B_{1} \rho_{0}^{3/2}+B_{1}
\sqrt{\rho_{0}}+B_{2}\rho_{0}+B_{2}-12 g_1 m^{-u}
\\\nonumber
&\times\big(H_{0}^2 \Upsilon^{2 q+2}\big)^{u+1},
\\\nonumber
P_{DE}&=\bigg[m^{-u} \Upsilon^{-3 (q+1)} \bigg(\frac{H_{0}^2
\Upsilon^{2 q+2}}{m^2}\bigg)^{-\frac{3 m}{8}}(2 u+1) m^u \bigg(3 a
\sqrt{\rho_{0}} \bigg\{-144 H_{0}^3 m \Upsilon^{3 q+3}
\\\nonumber
&-48 H_{0}^2 (4 H_{0}-m) (\Upsilon^{3 q+3}+m (3 m+8)\bigg)-36
H_{0}^3 (3 m+4)\Upsilon^{3 q+3} \bigg(\frac{H_{0}^2 \Upsilon^{2
q+2}}{m^2}\bigg)^{\frac{3 m}{8}}
\\\nonumber
&\times\bigg\{2 b+B_{1} \sqrt{\rho_{0}}+B_{2}\bigg\} -\sqrt{6} g_2 (3 m+4)
\sqrt{H_{0}^2 \Upsilon^{2 q+2}} \big(12 H_{0}^2\Upsilon^{3 q+3}+1\big)
\\\nonumber
&\times \bigg(\frac{H_{0}^2 \Upsilon^{2 q+2}}{m^2}\bigg)^{\frac{3
m}{8}}\bigg\} -24 g_1 H_{0} (3 m+4) \Upsilon^{q+1}
\bigg(\frac{H_{0}^2 \Upsilon^{2 q+2}}{m^2}\bigg)^{\frac{3 m}{8}}
\big(H_{0}^2 \Upsilon^{2 q+2}\big)^u
\\\nonumber
&\times \bigg\{-72 H_{0}^4 (u+1)\Upsilon^{4 q+4}+24 H_{0}^3 (u+1)
\Upsilon^{4 q+4} -4 H_{0} u (u+1) \Upsilon^{q+1}
\\\nonumber
&+36 H_{0}^4 \Upsilon^{4 q+4}\bigg\}\bigg)\bigg] \bigg[72 H_{0}^3 (3
m+4) (2 u+1)\bigg]^{-1}.
\end{align}
\begin{figure}\center
\epsfig{file=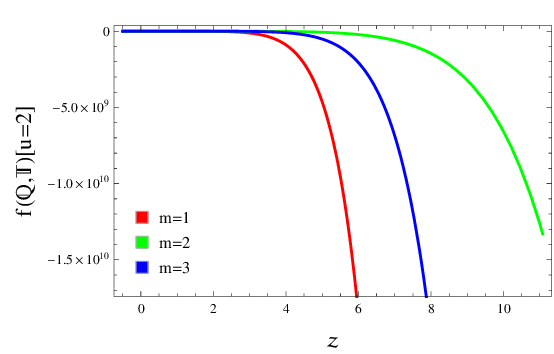,width=.5\linewidth}\epsfig{file=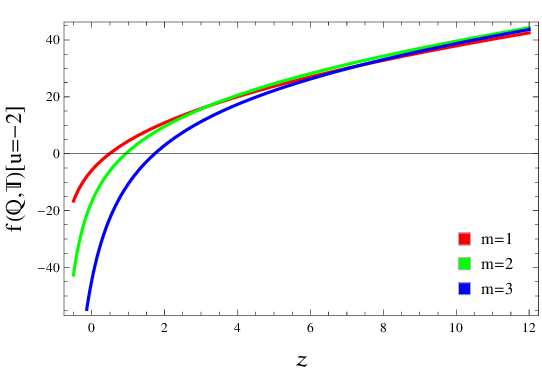,width=.46\linewidth}
\caption{Graph of the DE $f\mathbb{(Q,T)}$ model corresponding to $u
= \pm 2$ as a function of $z$.}
\end{figure}
\begin{figure}\center
\epsfig{file=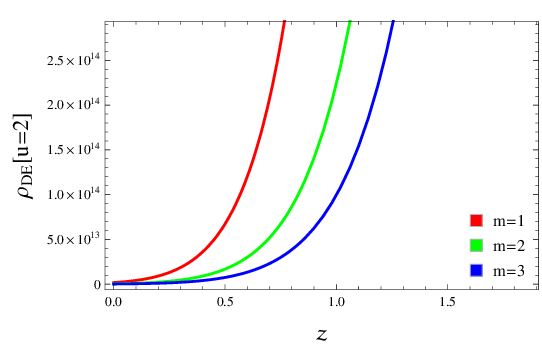,width=.5\linewidth}\epsfig{file=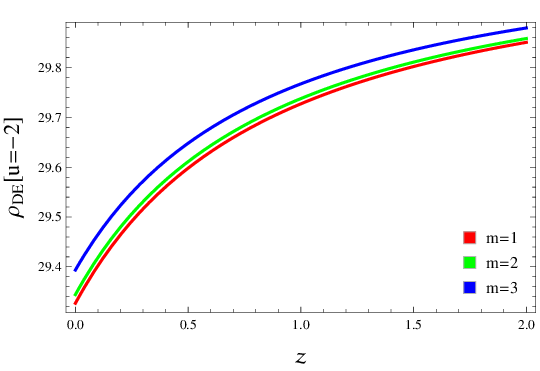,width=.46\linewidth}
\caption{Graphs of $\rho_{DE}$ $(u=\pm 2)$ versus $z$.}
\end{figure}
\begin{figure}\center
\epsfig{file=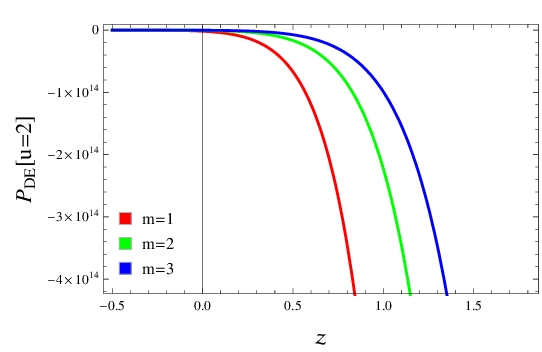,width=.5\linewidth}\epsfig{file=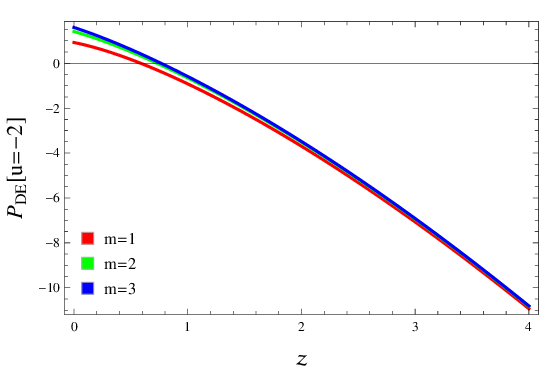,width=.46\linewidth}
\caption{Graphs of $P_{DE}$ $(u=\pm 2)$ against $z$.}
\end{figure}

\section{Cosmological Examination}

This section of the study delves into the cosmological development
of the universe across various epochs. We employ the reconstructed
PDE $f(\mathbb{Q, T})$ framework within a non-interacting context,
as elaborated in Eq.\eqref{52}. Additionally, the evolution of
critical cosmological parameters, including the EoS, the
$\omega_{DE}-\omega'_{DE}$ plane and the statefinder planes, is
illustrated. The stability of the framework is rigorously assessed
as well.

\subsection{The EoS Parameter}

The EoS parameter, denoted by $\omega = \frac{P_{DE}}{\rho_{DE}}$,
is essential for understanding both the phase of cosmic inflation
and the universe's ongoing expansion. We examine the conditions
required for a universe to accelerate, which occurs when $\omega$ is
less than $-\frac{1}{3}$. Specifically, $\omega = -1$ corresponds to
the cosmological constant, while $\omega = \frac{1}{3}$ and $\omega
= 0$ represent a universe dominated by radiation and matter,
respectively. Additionally, the phantom regime emerges when $\omega$
falls below -1. The EoS parameter is formulated as follows
\begin{equation}\label{53}
\omega_{DE}=
\frac{P_{eff}}{\rho_{eff}}=\frac{P_{DE}}{\rho_{DE}+\rho_{M}}.
\end{equation}
Equations \eqref{42}, \eqref{43}, and \eqref{43-a} are applied to
the above expression to calculate the relevant parameter as follows
\begin{align}\nonumber
\omega_{DE}&=2^{-\frac{3 m}{8}-u-\frac{7}{2}} 3^{-\frac{3
m}{8}-u-\frac{5}{2}} m^{-u} \bigg(\frac{H_{0}^2 \Upsilon^{2
q+2}}{m^2}\bigg)^{-\frac{3 m}{8}} \bigg[6^{u+\frac{1}{2}} (2 u+1)
m^u \sqrt{H_{0}^2 \Upsilon^{2 q+2}}
\\\nonumber
&\times\bigg\{a \sqrt{\rho_{0}} 6^{\frac{3 m}{8}} \big(-432 H_{0}^4
m \Upsilon^{4 q+4}-48 H_{0}^2 \Upsilon^{2 q+2} \big(12 H_{0}^2
\Upsilon^{2 q+2}-3 H_{0} m \Upsilon^{2 q+2}\big)
\\\nonumber
&+3 H_{0} m (3 m+8) \Upsilon^{q+1}\big)+H_{0}^4 6^{\frac{3 m}{8}+2}
(3 m+4) \Upsilon^{4 q+4} \bigg(\frac{H_{0}^2 \Upsilon^{2
q+2}}{m^2}\bigg)^{\frac{3 m}{8}} \bigg\{-2 b
\\\nonumber
&-B_{1} \sqrt{\rho_{0}}-B_{2}\bigg\}-g_2 6^{\frac{3 m}{8}+\frac{1}{2}} (3 m+4)
\sqrt{H_{0}^2 \Upsilon^{2 q+2}} \bigg\{-36 H_{0}^4 \Upsilon^{4 q+4}+6 H_{0}^2
\Upsilon^{2 q+2}
\\\nonumber
&\times \big(6 H_{0}^2 \Upsilon^{2 q+2} +2 H_{0}\Upsilon^{2
q+2}\big) +H_{0} \Upsilon^{q+1}\bigg\} \bigg(\frac{H_{0}^2
\Upsilon^{2 q+2}}{m^2}\bigg)^{\frac{3 m}{8}}\bigg\}\times2^{\frac{3
m}{8}+u+\frac{7}{2}} 3^{\frac{3 m}{8}+u+\frac{3}{2}}
\\\nonumber
& -g_1 (3 m+4)\bigg(\frac{H_{0}^2 \Upsilon^{2
q+2}}{m^2}\bigg)^{\frac{3 m}{8}} \big(H_{0}^2 \Upsilon^{2
q+2}\big)^{u+\frac{3}{2}} \big(-72 H_{0}^4 (u+1) \Upsilon^{4 q+4}
\\\nonumber
&+36 H_{0}^4 \Upsilon^{4 q+4}+24 H_{0}^3 (u+1) \Upsilon^{4 q+4}-4
H_{0}u (u+1) \Upsilon^{q+1}\big)\bigg]\bigg((3 m+4)
\\\nonumber
&\times (2 u+1) \big(H_{0}^2 \Upsilon^{2 q+2}\big)^{5/2} \bigg\{2 a \sqrt{\rho_{0}}
\bigg(\frac{H_{0}^2 \Upsilon^{2 q+2}}{m^2}\bigg)^{-\frac{3 m}{8}}+b+B_{1}
\sqrt{\rho_{0}}
\\\nonumber
&+B_{2}-12 g_1 m^{-u} \bigg(H_{0}^2 \Upsilon^{2 q+2}\bigg)^{u+1}\bigg\}.
\end{align}
Figure \textbf{6} illustrates the dynamics of the EoS parameter as a
function of $z$, demonstrating that both $u = 2$ and $u = -2$ align
with a phantom epoch throughout the present and ensuing stages of
cosmic evolution.
\begin{figure}\center
\epsfig{file=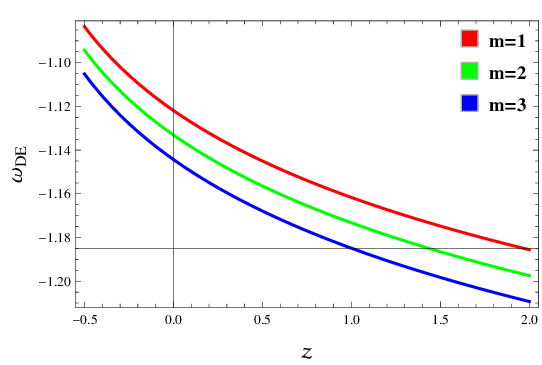,width=.46\linewidth}\epsfig{file=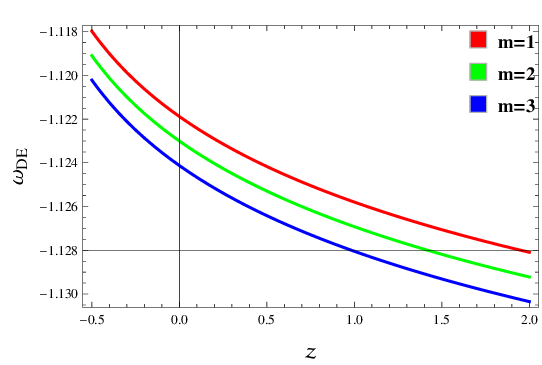,width=.46\linewidth}
\caption{Graph of the EoS parameter for $u = \pm 2$ versus $z$.}
\end{figure}

\subsection{$\omega_{DE}-\omega'_{DE}$ Plane}

Here, we examine the $\omega_{DE}-\omega'_{DE}$ plane
($\omega'_{DE}$ represents the evolution of $\omega_{DE}$, while the
prime denotes differentiation in terms of $\mathbb{Q}$). Caldwell
and Linder \cite{32} introduced a phase plane to analyze the cosmic
expansion within the quintessence DE model. Their findings showed
that the plane can be separated into two unique regions, thawing
$(\omega_{DE}<0, \omega'_{DE}>0 )$ and freezing $(\omega_{DE}
<0,~\omega'_{DE}<0 )$. In depicting the current model of cosmic
expansion, the freezing region indicates a more rapid rate of
acceleration as compared to the thawing region. Figure \textbf{7}
presents the cosmic trajectories in the $\omega_{DE} - \omega'_{DE}$
plane, illustrating the universe freezing phase under specific
values of the parameters $m$ and $u$. In this phase, the conditions
$\omega_{DE} < 0$ and $\omega'_{DE} < 0$ are satisfied, indicating
that the rate of cosmic expansion is accelerating. The expression
for $\omega'_{DE}$ can be found in Appendix \textbf{A}.
\begin{figure}\center
\epsfig{file=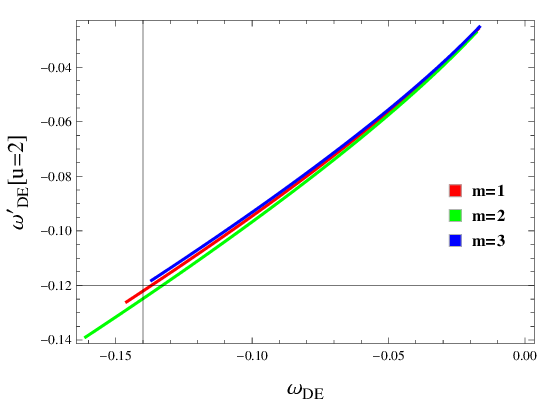,width=.44\linewidth}\epsfig{file=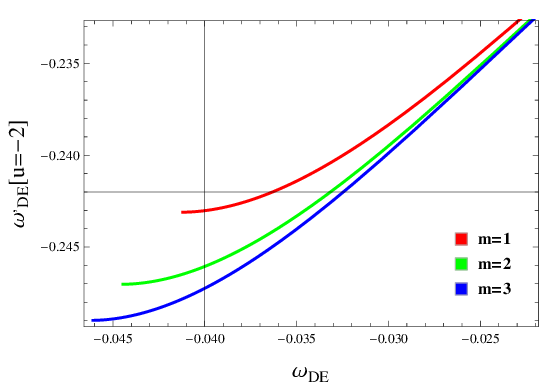,width=.46\linewidth}
\caption{$\omega_{DE}-\omega'_{DE}$ $(u=\pm 2)$ against $z$.}
\end{figure}

\subsection{$r-s$ Plane}

The deceleration and Hubble parameters provide a clear description
of the evolving cosmos. Nevertheless, numerous DE models have
identical parametric values at present. A set of unitless
cosmological parameters known as statefinders, derived from the
deceleration and Hubble parameters, is used to classify DE models,
as \cite{33-a}
\begin{equation}\label{64}
r=\frac{\dddot{a}}{a H^{3}}, \quad s=\frac{r-1}{3(q-\frac{1}{2})}.
\end{equation}
The parameter $r$ is defined with respect to the deceleration
parameter as follows
\begin{equation}\label{64-a}
r=2q^{2}+q-\acute{q}.
\end{equation}
The parameter $s$ controls the speed of cosmic acceleration, whereas
the parameter $r$ reflects variations from a strict \emph{power-law}
expansion. The dynamics of this diagnostic pair is crucial for
illustrating different cosmic phases and distinguishing between
various DE models. When $(r, s) = (1, 0)$, it signifies the cold
dark matter (CDM) regime, whereas $(r, s) = (1, 1)$ corresponds to
the $\Lambda$CDM model. The region where $r < 1$ and $s > 0$
includes both phantom and non-phantom phases, while the Chaplygin
gas (CG) model is defined by $r > 1$ and $s < 0$. Figure \textbf{8}
illustrates the $r-s$ phase plane, showing $r > 1$ and $s < 0$,
which verifies the presence of the CG model. This indicates that our
model effectively captures the universe's transition to an
accelerating expansion, consistent with observational evidence of
cosmic acceleration.
\begin{figure}\center
\epsfig{file=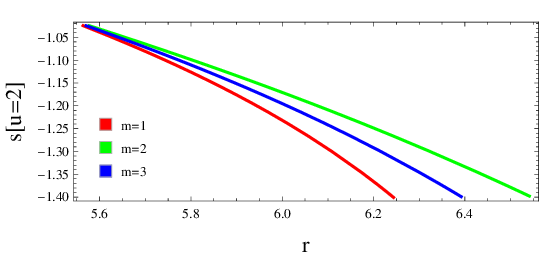,width=.5\linewidth}\epsfig{file=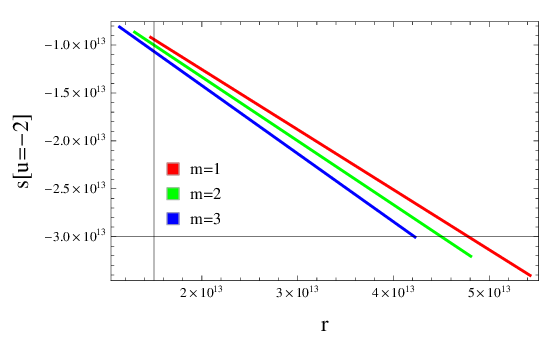,width=.46\linewidth}
\caption{Graph of the $r-s$ plane for $u = \pm 2$ as a function of
$z$.}
\end{figure}

\subsection{$\nu_{s}^{2}$ Parameter}

The stability of the DE framework can be directly evaluated using
perturbation theory by analyzing the sign of the $\nu_{s}^{2}$
parameter. In this context, we examine the $\nu_{s}^{2}$ to evaluate
the stability of the PDE $f\mathbb{(Q,T)}$ approach, as outlined
\cite{62-a}
\begin{equation}\label{55}
\nu_{s}^{2}=\frac{P'_{eff}}{\rho'_{eff}}.
\end{equation}
A positive value indicates that the configuration is stable, while a
negative value points to instability within the associated
framework. By substituting the required expressions into the
equation for the reconstructed framework, we obtain the
$\nu_{s}^{2}$ parameter, which is elaborated upon in Appendix
\textbf{B}. As shown in Figure \textbf{9}, all tested values of $m$
and $u$ demonstrate positive behavior, validating the stability of
the reconstructed $f(\mathbb{Q,T})$ DE framework during the
universe's evolution. This stability across varying parameters
highlights the framework's robustness and its capability to
represent a consistent cosmic evolution.
\begin{figure}\center
\epsfig{file=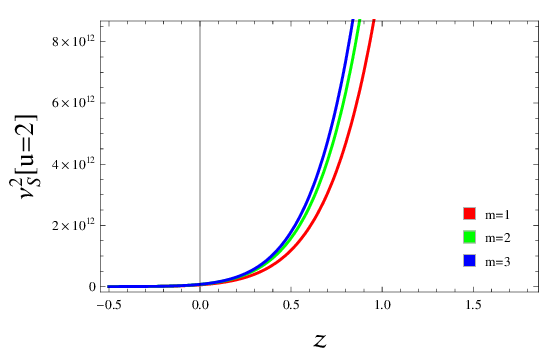,width=.46\linewidth}\epsfig{file=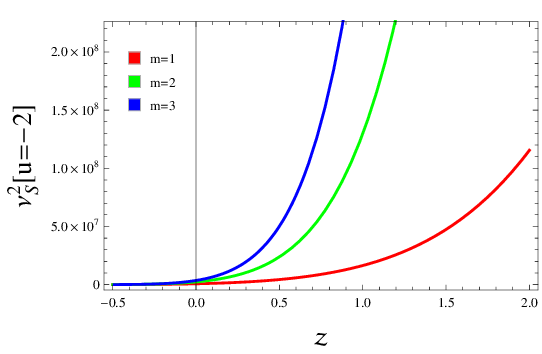,width=.46\linewidth}
\caption{Plot of $\nu_{s}^{2}$ $(u=\pm 2)$ versus $z$.}
\end{figure}

\subsection{Jerk, Snap, Lerk and Maxout parameters}

One of the central challenges in modern cosmology is distinguishing
between a cosmological constant and alternative DE models as
explanations for the universe accelerated expansion. To develop
robust, model-independent diagnostic tools, it becomes essential to
examine cosmographic parameters that extend beyond the deceleration
parameter. These include the jerk, snap, lerk, and maxout
parameters, which provide a higher-order description of cosmic
kinematics. These cosmographic quantities are derived from
successive time derivatives of the scale factor and serve to refine
our understanding of deviations from standard cosmological
scenarios. Specifically, they are defined as:
\begin{eqnarray}\nonumber
J&=&\frac{1}{a H^3}\frac{d^{3}a}{dt^{3}}, \quad S=\frac{1}{a
H^4}\frac{d^{4}a}{dt^{4}},\\ L&=&\frac{1}{a
H^5}\frac{d^{5}a}{dt^{5}}, \quad M=\frac{1}{a
H^6}\frac{d^{6}a}{dt^{6}},
\end{eqnarray}
where $J, S, L, M$ correspond to the jerk, snap, lerk and maxout
parameters, respectively. These higher order descriptors offer a
detailed perspective on the universe expansion dynamics, extending
the traditional cosmographic series composed of the Hubble and
deceleration parameters. While they emerge from the scale factor's
derivatives, they also relate closely to higher order derivatives of
position or displacement in kinematic terms, thus playing a vital
role in the dynamical analysis of cosmic evolution. Notably, these
parameters are not mutually independent but are interconnected
through well defined mathematical relationships. Figure \textbf{10}
illustrates the evolution of these cosmographic indicators,
revealing a positive trend consistent with ongoing cosmic
acceleration. Moreover, their behavior mirrors key features expected
from the $\Lambda$CDM model, lending further credibility to its
framework within observational limits.
\begin{figure}\center
\epsfig{file=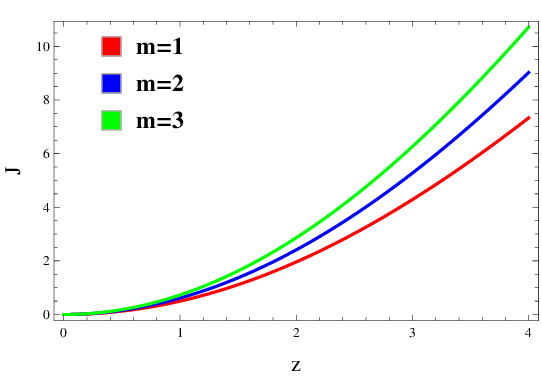,width=.46\linewidth}\epsfig{file=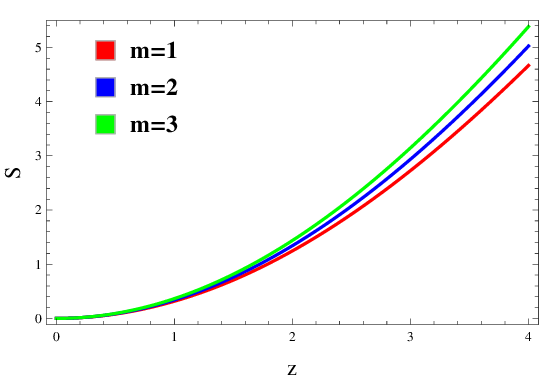,width=.46\linewidth}
\epsfig{file=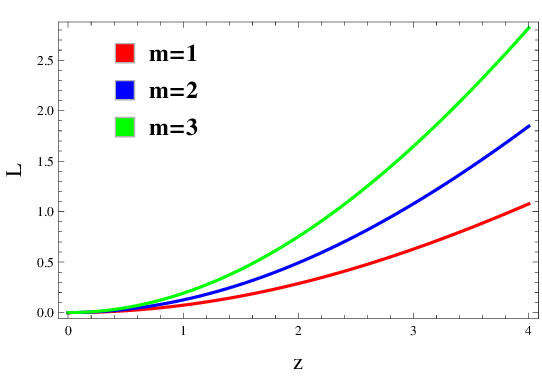,width=.46\linewidth}\epsfig{file=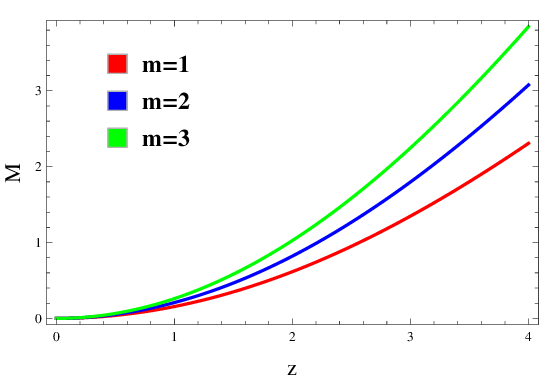,width=.46\linewidth}
\caption{Graphs of Jerk, Snap, Lerk and Maxout against $z$.}
\end{figure}

\section{Summary}

This study explores the features of the PDE model for a \emph{power
law scale factor} of the flat FRW universe. We have applied a
reconstruction approach through the correspondence principle within
$f(\mathbb{Q,T})$ theory. The reconstructed PDE $f\mathbb{(Q,T)}$
model's cosmic evolutionary paradigm has been assessed using
cosmological parameters. To assess the effects of $\mathbb{Q}$ and
$\mathbb{T}$ within the PDE framework, we have a selected model of
this theory. We have reconstructed $f\mathbb{(Q,T)}$ models based on
two PDE parameter values $u=\pm2$. We have then examined the
equation of state parameter, $\omega_{DE}-\omega'_{DE}$, along with
the statefinder planes. Lastly, we assessed the stability of the
derived paradigm by analyzing the $\nu_{s}^{2}$ parameter. Here is a
summary of the main results.
\begin{itemize}
\item The reconstructed PDE $f\mathbb{(Q,T)}$ model exhibits a decreasing and
increasing trend in $z$ when $u=2$ and $u=-2$, respectively that satisfy the
monotonic property of pilgrim parameters (Figure \textbf{3}).
\item The $\rho_{DE}$ displays positive behavior, marked by an
increase, whereas the $P_{DE}$ shows negative behavior. These patterns align closely
to the properties of DE (Figures \textbf{4} and \textbf{5}).
\item
The EoS parameter reveals a phantom phase in the non interacting
scenario (Figure \textbf{6}).
\item
The developmental path in the $\omega_{DE}-\omega'_{DE}$ plane
indicates a freezing region across every value of $m$ and $u$
(Figure \textbf{7}). This finding supports the conclusion that the
non interacting DE $f(\mathbb{Q,T})$ gravity predicts a universe
with rapid acceleration.
\item The $r-s$ plane illustrates the consistency of the
CG framework, where $r$ and $s$ fulfill the specified requirements for the model
(Figure \textbf{8}).
\item Our analysis shows that the $\nu_{s}^{2}$ parameter is positive, indicating
that the PDE $f(\mathbb{Q,T})$ gravity stable across all variations of $m$ and $u$
(Figure \textbf{9}).
\end{itemize}

The realization of inflation within the PDE model in
$f(\mathbb{Q,T})$ gravity is theoretically feasible, as this
modified framework supports alternative cosmological behavior beyond
GR. Unlike standard inflation, the geometric modifications in
$f(\mathbb{Q,T})$ gravity can naturally induce accelerated
expansion. Phase plane analysis through
$\omega_{\text{DE}}-\omega'_{\text{DE}}$ and statefinder planes,
reveals freezing behavior and the CG model, respectively, where
rapid acceleration stabilizes over time, a feature consistent with
both early cosmic inflation and late time acceleration phases.
Notably, the model's phantom regime, $\omega_{\text{DE}} < -1$,
aligns with conditions conducive to inflation and the stability of
the model is confirmed by positive $\nu_s^2$, indicating its ability
to maintain stable configurations across cosmic phases. This is
essential, enabling controlled expansion without disruptive
instabilities, a key criterion for successful inflationary behavior.

In conclusion, the results of our study align with the latest
observational data, as detailed below \cite{34-a}
\begin{eqnarray}\nonumber
\omega_{DE}&=&-1.023^{+0.091}_{-0.096}\quad(\text{Planck
TT+LowP+ext}),\\\nonumber
\omega_{DE}&=&-1.006^{+0.085}_{-0.091}\quad(\text{Planck
TT+LowP+lensing+ext}),\\\nonumber
\omega_{DE}&=&-1.019^{+0.075}_{-0.080}\quad (\text{Planck TT, TE,
EE+LowP+ext}).
\end{eqnarray}
We have derived our results using a range of observational methods
with a 95$\%$ degree of certainty, ensuring that the statefinder
parameters employed for cosmic diagnostics in our framework align
with configurational constraints and cosmic boundaries established
in previous studies \cite{35}. While prior works in $f(T)$,
$f\mathbb{(Q)}$ and $f\mathbb{(G, T)}$ theories of gravity have
explored similar themes, our study builds upon these efforts to
address gaps and offer new insights. Sharif and Rani \cite{6-d}
investigated $f(T)$ gravity with an interacting PDE model, revealing
a phantom era for certain pilgrim parameter ranges, freezing regions
in the $\omega_{DE}-\omega'_{DE}$ plane and a $\Lambda$CDM limit in
the statefinder plane. Myrzakulov et al. \cite{6-g} reconstructed
$f(\mathbb{Q})$ gravity using a power law cosmology to examine PDE
in an interacting scenario but lacked emphasis on stability
analysis. Sharif and Saba \cite{62} analyzed $f\mathbb{(G, T)}$
gravity via a correspondence method, identifying accelerated
expansion and phantom like behavior. In contrast, our study employs
$f(\mathbb{Q,T})$ gravity, where non-metricity plays a central role,
setting it apart from the $f(T)$ and $f(\mathbb{G,T})$ approaches.
We have presented a comprehensive, non-interacting PDE framework,
supported by detailed phase plane analysis, revealing a freezing
region in the $\omega_{DE} - \omega'_{DE}$ plane and CG behavior in
the $r - s$ plane, consistent with observational data. Additionally,
our model emphasizes stability through the $\nu_s^2$, providing a
robust analysis often overlooked in the earlier studies. The PDE
aims to provide insights into the dynamics of the universe and by
applying $f(\mathbb{Q,T})$ gravity, we offer a promising avenue for
addressing key cosmological issues and understanding its long term
evolution. Our findings align strongly with recent observational
evidence, making this work a significant contribution that
complements and extends previous studies in modified gravity
frameworks.

Our approach explored key cosmological parameters including the EoS
$\omega_{DE}$, phase space diagnostics, and the stability parameter
$v_s^2$. A crucial aspect of our formulation lies in the geometrical
foundation of the theory specifically, the non-metricity scalar
$\mathbb{Q}= 6H^2$, which governs the gravitational action. Through
the redshift evolution of $H(z)$, we indirectly traced the behavior
of $Q(z)$, which plays a pivotal role in shaping the dynamical
features of the universe in this model. The reconstructed
$f\mathbb{(Q,T)}$ function exhibited consistent behavior with
cosmological expectations: the derived $f\mathbb{(Q,T)}$ decreases
with redshift, representing a universe transitioning from a
decelerated expansion phase in the past to accelerated expansion at
present. This dynamic aligns with current observational constraints,
including Planck data that favor a DE EoS near $\omega_{DE} \approx
-1$. Furthermore, the geometrical quantity $\mathbb{Q}$ facilitated
a natural phantom regime without invoking exotic fields, confirming
that the late time acceleration and phantom behavior are
geometrically driven. Therefore, our results demonstrate that the
geometrical evolution of non metricity within the $f\mathbb{(Q,T)}$
framework not only supports a stable cosmic scenario but also aligns
with contemporary observational data, validating the theoretical
robustness of the proposed model.

We have compared the predictions of our reconstructed
$f\mathbb{(Q,T)}$ gravity model with recent cosmological
observations provided by the DE spectroscopic instrument (DESI). Our
model predicts a phantom-like EoS parameter, where $\omega_{DE} <
-1$, a feature that remains well within the observational bounds
reported by DESI. Specifically, DESI 2024 data release constrains
the DE EoS to $\omega_{DE} = -1.01 \pm 0.06$, which is consistent
with the phantom regime predicted in our framework. Moreover, the
redshift-dependent behavior of the Hubble parameter derived from our
power-law cosmology aligns closely with the DESI measurements across
the range $0.1 < z < 2$, where DESI has provided high-precision
expansion rate data using baryon acoustic oscillations and
redshift-space distortions. Our model naturally exhibits a smooth
transition from early-time deceleration to late-time acceleration
around $z \approx 0.7$, in excellent agreement with DESI
reconstructed expansion history. These correspondences validate the
physical viability of the $f\mathbb{(Q,T)}$ gravity model and
confirm that the geometrically induced cosmic evolution is not only
mathematically consistent but also in strong agreement with the most
recent cosmological observations.

\section*{Appendix A: Calculation of $\omega'_{DE}$}
\renewcommand{\theequation}{A\arabic{equation}}
\setcounter{equation}{0}
\begin{align}\nonumber
\omega'_{DE}&=\bigg[2^{-\frac{3 m}{8}-u-\frac{7}{2}} 3^{-\frac{3
m}{8}-u-\frac{5}{2}} m^{-u} \bigg(\frac{H_{0}^2 \Upsilon^{2
q+2}}{m^2}\bigg)^{-\frac{3 m}{8}} \bigg\{-2^{\frac{3
m}{8}+u+\frac{7}{2}} 3^{\frac{3 m}{8}+u+\frac{3}{2}}
\\\nonumber
&\times(3 m+4)\big(H_{0}^2 \Upsilon^{2 q+2}\big)^{u+\frac{3}{2}} \bigg\{12 H_{0}^2
\Upsilon^{2 q+2}-12 H_{0}^2 (u+1) \Upsilon^{2 q+2}+4 H_{0} (u+1)
\\\nonumber
&\times \Upsilon^{2 q+2}\bigg\} g_{1} \bigg(\frac{H_{0}^2
\Upsilon^{2 q+2}}{m^2}\bigg)^{\frac{3 m}{8}}-2^{\frac{3
m}{8}+u+\frac{5}{2}} 3^{\frac{3 m}{8}+u+\frac{1}{2}}
\bigg(u+\frac{3}{2}\bigg) \big(H_{0}^2 \Upsilon^{2
q+2}\big)^{u+\frac{1}{2}}
\\\nonumber
&\times(3 m+4) \bigg(-4 H_{0} u (u+1) \Upsilon^{q+1}+36 H_{0}^4
\Upsilon^{4 q+4} -72 H_{0}^4 (u+1) \Upsilon^{4 q+4}
\\\nonumber
&+24 H_{0}^3 (u+1) \Upsilon^{4 q+4}\bigg) g_1 \bigg(\frac{H_{0}^2
\Upsilon^{2 q+2}}{m^2}\bigg)^{\frac{3 m}{8}} -\bigg[2^{\frac{3
m}{8}+u-\frac{1}{2}} 3^{\frac{3 m}{8}+u+\frac{3}{2}} (3 m+4)
\\\nonumber
&\times \bigg(H_{0}^2 \Upsilon^{2 q+2}\bigg)^{u+\frac{3}{2}}
\bigg(\frac{H_{0}^2 \Upsilon^{2 q+2}}{m^2}\bigg)^{\frac{3 m}{8}-1}
\bigg\{-4 H_{0} u (u+1) \Upsilon^{q+1} +36 H_{0}^4 \Upsilon^{4 q+4}
\\\nonumber
& -72 H_{0}^4 (u+1) \Upsilon^{4 q+4}+24 H_{0}^3 (u+1) \Upsilon^{4
q+4}\bigg\} g_1\bigg]\bigg[m\bigg]^{-1}+\bigg[2^{u-\frac{3}{2}}
3^{u-\frac{1}{2}} m^u
\\\nonumber
&\times(2 u+1) \bigg\{6^{\frac{3 m}{8}+2} \bigg\{-2 b -B_{2}-B_{1} \sqrt{-H_{0}
\Upsilon^{2 q+2}}\bigg\} H_{0}^4 (3 m+4) \Upsilon^{4 q+4}
\\\nonumber
&\times\bigg(\frac{H_{0}^2 \Upsilon^{2 q+2}}{m^2}\bigg)^{\frac{3
m}{8}} -6^{\frac{3 m}{8}+\frac{1}{2}} (3 m+4) \sqrt{H_{0}^2
\Upsilon^{2 q+2}} \big(H_{0} \Upsilon^{q+1}+6 H_{0}^2 \big(6 H_{0}^2
\Upsilon^{2 q+2}
\\\nonumber
&+2 H_{0} \Upsilon^{2 q+2}\big) \Upsilon^{2 q+2}-36 H_{0}^4
\Upsilon^{4 q+4}\big) g_2 \bigg(\frac{H_{0}^2 \Upsilon^{2
q+2}}{m^2}\bigg)^{\frac{3 m}{8}}+6^{\frac{3 m}{8}} a \sqrt{\rho_{0}}
\big(3 H_{0} m
\\\nonumber
&\times(3 m+8) \Upsilon^{q+1} -48 H_{0}^2 \big(12 H_{0}^2
\Upsilon^{2 q+2}-3 H_{0} m \Upsilon^{2 q+2}\big) \Upsilon^{2
q+2}-432 H_{0}^4
\\\nonumber
&\times m \Upsilon^{4 q+4}\big)\bigg\}\bigg] \bigg[\sqrt{H_{0}^2
\Upsilon^{2 q+2}}\bigg]^{-1}+6^{u+\frac{1}{2}} m^u (2 u+1)
\sqrt{H_{0}^2 \Upsilon^{2 q+2}} \bigg\{2^{\frac{3 m}{8}+2}
3^{\frac{3 m}{8}+1}
\\\nonumber
&\times\bigg\{-2 b-B_{2}-B_{1} \sqrt{-H_{0} \Upsilon^{2 q+2}}\bigg\} H_{0}^2 (3 m+4)
\Upsilon^{2 q+2} \bigg(\frac{H_{0}^2 \Upsilon^{2 q+2}}{m^2}\bigg)^{\frac{3 m}{8}}
\\\nonumber
&-6^{\frac{3 m}{8}+\frac{1}{2}} (3 m+4) \sqrt{H_{0}^2 \Upsilon^{2
q+2}} \bigg(6 H_{0}^2 \Upsilon^{2 q+2}+2 H_{0} \Upsilon^{2
q+2}\bigg) g_2 \bigg(\frac{H_{0}^2 \Upsilon^{2
q+2}}{m^2}\bigg)^{\frac{3 m}{8}}
\\\nonumber
&+\bigg[2^{\frac{3 m}{8}-2} 3^{\frac{3 m}{8}+2} \bigg(-2 b-B_{2}-B_{1} \sqrt{-H_{0}
\Upsilon^{2 q+2}}\bigg) H_{0}^4 (3 m+4) \Upsilon^{4 q+4}
\\\nonumber
&\times\bigg\{\frac{H_{0}^2 \Upsilon^{2 q+2}}{m^2}\bigg\}^{\frac{3
m}{8}-1}\bigg]\bigg[m\bigg]^{-1} +6^{\frac{3 m}{8}} a
\sqrt{\rho_{0}} \bigg\{-96 H_{0}^2 \Upsilon^{2 q+2}-72 H_{0}^2 m
\Upsilon^{2 q+2}
\\\nonumber
&-8 \big(12 H_{0}^2 \Upsilon^{2 q+2}-3 H_{0} m \Upsilon^{2
q+2}\big)\bigg\} -\bigg[2^{\frac{3 m}{8}-\frac{3}{2}} 3^{\frac{3
m}{8}-\frac{1}{2}} (3 m+4) \bigg(\frac{H_{0}^2 \Upsilon^{2
q+2}}{m^2}\bigg)^{\frac{3 m}{8}}
\\\nonumber
&\times \big(H_{0} \Upsilon^{q+1} +6 H_{0}^2 \big(6 H_{0}^2
\Upsilon^{2 q+2}+2 H_{0} \Upsilon^{2 q+2}\big) \Upsilon^{2 q+2}-36
H_{0}^4 \Upsilon^{4 q+4}\big) g_2\bigg]
\\\nonumber
&\times\bigg[\sqrt{H_{0}^2 \Upsilon^{2 q+2}}\bigg]^{-1}
-\bigg[2^{\frac{3 m}{8}-\frac{7}{2}} 3^{\frac{3 m}{8}+\frac{1}{2}}
(3 m+4) \sqrt{H_{0}^2 \Upsilon^{2 q+2}} \bigg(\frac{H_{0}^2
\Upsilon^{2 q+2}}{m^2}\bigg)^{\frac{3 m}{8}-1}
\\\nonumber
&\times \bigg\{H_{0} \Upsilon^{q+1} +6 H_{0}^2 \big(6 H_{0}^2
\Upsilon^{2 q+2}+2 H_{0} \Upsilon^{2 q+2}\big) \Upsilon^{2 q+2}-36
H_{0}^4 \Upsilon^{4 q+4}\bigg\} g_2\bigg]
\\\nonumber
&\times\bigg[m\bigg]\bigg\}\bigg\}\bigg]\bigg[(3 m+4) (2 u+1) \big(H_{0}^2
\Upsilon^{2 q+2}\big)^{5/2} \bigg\{\rho_{0} +\frac{1}{2} \bigg(4 a \sqrt{\rho_{0}}
\bigg(\frac{H_{0}^2 \Upsilon^{2 q+2}}{m^2}\bigg)^{-\frac{3 m}{8}}
\\\nonumber
&+3 b+B_{2}+2 b \rho_{0}+\sqrt{a\rho_{0}} (2 \rho_{0}+1) -24 m^{-u}
\big(H_{0}^2 \Upsilon^{2 q+2}\big)^{u+1} g_1
\\\nonumber
&+B_{1} \sqrt{-H_{0} \Upsilon^{2 q+2}}\bigg)\bigg\}\bigg]^{-1}
-\bigg[2^{-\frac{3m}{8} -u-\frac{15}{2}} 3^{-\frac{3 m}{8}-u-\frac{5}{2}} m^{-u-1}
\bigg(\frac{H_{0}^2 \Upsilon^{2 q+2}}{m^2}\bigg)^{-\frac{3 m}{8}-1}
\\\nonumber
&\times \big(6^{u+\frac{1}{2}} m^u (2 u+1) \sqrt{H_{0}^2 \Upsilon^{2 q+2}}
\big(6^{\frac{3 m}{8}+2} \bigg\{-2 b-B_{2} -B_{1} \sqrt{-H_{0} \Upsilon^{2
q+2}}\bigg\}
\\\nonumber
&\times H_{0}^4 (3 m+4) \Upsilon^{4 q+4} \bigg(\frac{H_{0}^2
\Upsilon^{2 q+2}}{m^2}\bigg)^{\frac{3 m}{8}}-6^{\frac{3
m}{8}+\frac{1}{2}} (3 m+4) \sqrt{H_{0}^2 \Upsilon^{2 q+2}}
\bigg(H_{0} \Upsilon^{q+1}
\\\nonumber
&+6 H_{0}^2 \big(6 H_{0}^2 \Upsilon^{2 q+2}+2 H_{0} \Upsilon^{2
q+2}\big) \Upsilon^{2 q+2}-36 H_{0}^4 \Upsilon^{4 q+4}\bigg) g_2
\bigg(\frac{H_{0}^2 \Upsilon^{2 q+2}}{m^2}\bigg)^{\frac{3 m}{8}}
\\\nonumber
& +6^{\frac{3 m}{8}}\sqrt{\rho_{0}} \big(3 H_{0} m (3 m+8) \Upsilon^{q+1}-48 H_{0}^2
\big(12 H_{0}^2 \Upsilon^{2 q+2}-3 H_{0} m \Upsilon^{2 q+2}\big) \Upsilon^{2 q+2}
\\\nonumber
&-432 H_{0}^4 m \Upsilon^{4 q+4}\big)\big) -2^{\frac{3
m}{8}+u+\frac{7}{2}} 3^{\frac{3 m}{8}+u+\frac{3}{2}} (3 m+4)
\big(H_{0}^2 \Upsilon^{2 q+2}\big)^{u+\frac{3}{2}}
\bigg(\frac{H_{0}^2 \Upsilon^{2 q+2}}{m^2}\bigg)^{\frac{3 m}{8}}
\\\nonumber
&\times\big(-4 H_{0} u (u+1) \Upsilon^{q+1} +36 H_{0}^4 \Upsilon^{4
q+4}-72 H_{0}^4 (u+1) \Upsilon^{4 q+4}+24 H_{0}^3 (u+1)
\\\nonumber
&\times \Upsilon^{4 q+4}\big) g_1\big)\bigg]\bigg[(3 m+4) (2 u+1) \big(H_{0}^2
\Upsilon^{2 q+2}\big)^{5/2} \bigg(\rho_{0}+\frac{1}{2} \bigg(4 a \sqrt{\rho_{0}}
\bigg(\frac{H_{0}^2 \Upsilon^{2 q+2}}{m^2}\bigg)^{-\frac{3 m}{8}}
\\\nonumber
&+3 b+B_{2}+2 b+\sqrt{\rho_{0}} (2 +1) -24 m^{-u} \bigg(H_{0}^2 \Upsilon^{2
q+2}\bigg)^{u+1} g_1
\\\nonumber
&+B_{1} \sqrt{-H_{0} \Upsilon^{2 q+2}}\bigg)\bigg)\bigg]^{-1} -\bigg[5\ 2^{-\frac{3
m}{8}-u-\frac{11}{2}} 3^{-\frac{3 m}{8}-u-\frac{7}{2}} m^{-u} \bigg(\frac{H_{0}^2
\Upsilon^{2 q+2}}{m^2}\bigg)^{-\frac{3 m}{8}}
\\\nonumber
&\times\big(6^{u+\frac{1}{2}} m^u (2 u+1) \sqrt{H_{0}^2 \Upsilon^{2 q+2}}
\bigg(6^{\frac{3 m}{8}+2} \bigg\{-2 b-B_{2}-B_{1} \sqrt{-H_{0} \Upsilon^{2
q+2}}\bigg\}
\\\nonumber
&\times H_{0}^4 (3 m+4)\Upsilon^{4 q+4} \bigg(\frac{H_{0}^2
\Upsilon^{2 q+2}}{m^2}\bigg)^{\frac{3 m}{8}}-6^{\frac{3
m}{8}+\frac{1}{2}} (3 m+4) \sqrt{H_{0}^2 \Upsilon^{2 q+2}}
\big(H_{0} \Upsilon^{q+1}
\\\nonumber
&+6 H_{0}^2 \big(6 H_{0}^2 \Upsilon^{2 q+2}+2 H_{0} \Upsilon^{2
q+2}\big) \Upsilon^{2 q+2}-36 H_{0}^4 \Upsilon^{4 q+4}\big) g_2
\bigg(\frac{H_{0}^2 \Upsilon^{2 q+2}}{m^2}\bigg)^{\frac{3 m}{8}}
\\\nonumber
&+6^{\frac{3 m}{8}}\sqrt{\rho_{0}} \bigg\{3 H_{0} m (3 m+8) \Upsilon^{q+1}-48 H_{0}^2
\big(12 H_{0}^2 \Upsilon^{2 q+2}-3 H_{0} m \Upsilon^{2 q+2}\big) \Upsilon^{2 q+2}
\\\nonumber
&-432 H_{0}^4 m \Upsilon^{4 q+4}\bigg\}\bigg)-2^{\frac{3
m}{8}+u+\frac{7}{2}} 3^{\frac{3 m}{8}+u+\frac{3}{2}} (3 m+4)
\bigg(H_{0}^2 \Upsilon^{2 q+2}\bigg)^{u+\frac{3}{2}}
\\\nonumber
&\times\bigg(\frac{H_{0}^2 \Upsilon^{2 q+2}}{m^2}\bigg)^{\frac{3
m}{8}} \bigg(-4 H_{0} u (u+1) \Upsilon^{q+1} +36 H_{0}^4
(\Upsilon+1)^{4 q+4}-72 H_{0}^4 (u+1)
\\\nonumber
& \times\Upsilon^{4 q+4} +24 H_{0}^3 (u+1) \Upsilon^{4 q+4}\bigg)
g_1\bigg)\bigg]\bigg[(3 m+4) (2 u+1) \big(H_{0}^2 \Upsilon^{2 q+2}\big)^{7/2}
\bigg(\rho_{0}
\\\nonumber
&+\frac{1}{2} \bigg(4\sqrt{\rho_{0}} \bigg(\frac{H_{0}^2 \Upsilon^{2
q+2}}{m^2}\bigg)^{-\frac{3 m}{8}}+3 b+B_{2}+2 b \rho_{0}+\sqrt{\rho_{0}} (2
\rho_{0}+1)-24 m^{-u}
\\\nonumber
&\times \big(H_{0}^2\Upsilon^{2 q+2}\big)^{u+1} g_1 +B_{1} \sqrt{-H_{0} \Upsilon^{2
q+2}}\bigg)\bigg)\bigg]^{-1} -\bigg[2^{-\frac{3 m}{8}-u-\frac{9}{2}} 3^{-\frac{3
m}{8}-u-\frac{5}{2}} m^{-u}
\\\nonumber
&\times \bigg(\frac{H_{0}^2 \Upsilon^{2 q+2}}{m^2}\bigg)^{-\frac{3
m}{8}}\bigg\{-\frac{\sqrt{\rho_{0}} \big(\frac{H_{0}^2 \Upsilon^{2
q+2}}{m^2}\big)^{-\frac{3 m}{8}-1}}{4 m}-4 m^{-u} (u+1) \bigg(H_{0}^2 \Upsilon^{2
q+2}\bigg) g_1\bigg\}
\\\nonumber
&\times \bigg(6^{u+\frac{1}{2}} m^u (2 u+1) \sqrt{H_{0}^2 \Upsilon^{2 q+2}}
\bigg(6^{\frac{3 m}{8}+2} \bigg(-2 b-B_{2}-B_{1} \sqrt{-H_{0} \Upsilon^{2 q+2}}\bigg)
\\\nonumber
&\times H_{0}^4 (3 m+4) \Upsilon^{4 q+4} \bigg(\frac{H_{0}^2 \Upsilon^{2
q+2}}{m^2}\bigg)^{\frac{3 m}{8}}-6^{\frac{3 m}{8}+\frac{1}{2}} (3 m+4) \sqrt{H_{0}^2
\Upsilon^{2 q+2}}
\\\nonumber
&\times \big(H_{0} \Upsilon^{q+1}+6 H_{0}^2 \big(6 H_{0}^2 \Upsilon^{2 q+2}+2 H_{0}
\Upsilon^{2 q+2}\big) \Upsilon^{2 q+2}-36 H_{0}^4 \Upsilon^{4 q+4}\big)
\\\nonumber
&\times g_2 \bigg(\frac{H_{0}^2 \Upsilon^{2 q+2}}{m^2}\bigg)^{\frac{3
m}{8}}+6^{\frac{3 m}{8}}\sqrt{\rho_{0}} \big(3 H_{0} m (3 m+8) \Upsilon^{q+1}-48
H_{0}^2 \big(12 H_{0}^2 \Upsilon^{2 q+2}
\\\nonumber
&-3 H_{0} m \Upsilon^{2 q+2}\big) \Upsilon^{2 q+2} -432 H_{0}^4 m
\Upsilon^{4 q+4}\big)\bigg)-2^{\frac{3 m}{8}+u+\frac{7}{2}}
3^{\frac{3 m}{8}+u+\frac{3}{2}} (3 m+4)
\\\nonumber
&\times \big(H_{0}^2 \Upsilon^{2 q+2}\big)^{u+\frac{3}{2}}
\bigg(\frac{H_{0}^2 \Upsilon^{2 q+2}}{m^2}\bigg)^{\frac{3 m}{8}}
\bigg(-4 H_{0} u (u+1) \Upsilon^{q+1} +36 H_{0}^4 \Upsilon^{4 q+4}
\\\nonumber
&-72 H_{0}^4 (u+1) \Upsilon^{4 q+4} +24 H_{0}^3 (u+1) \Upsilon^{4 q+4}\bigg)
g_1\bigg)\bigg]\bigg[(2 u+1) \big(H_{0}^2 \Upsilon^{2 q+2}\big)^{5/2} \bigg\{\rho_{0}
\\\nonumber
&+\frac{1}{2} \bigg(4\sqrt{\rho_{0}} \bigg(\frac{H_{0}^2 \Upsilon^{2
q+2}}{m^2}\bigg)^{-\frac{3 m}{8}} +3 b+B_{2}+2 b+\sqrt{\rho_{0}} (2 \rho_{0}+1)-24
m^{-u}
\\\label{61}
&\times\big(H_{0}^2 \Upsilon^{2 q+2}\big)^{u+1} g_1 +B_{1} \sqrt{-H_{0} \Upsilon^{2
q+2}}\bigg)\bigg\}\bigg]^{-1}.
\end{align}

\section*{Appendix B: Determination of $\nu_{s}^{2}$ }
\renewcommand{\theequation}{B\arabic{equation}}
\setcounter{equation}{0}
\begin{align}\nonumber
\nu_{s}^{2}&=\bigg [2^{-\frac{3 m}{8}-u-\frac{7}{2}} 3^{-\frac{3
m}{8}-u-\frac{5}{2}} m^{-u} \bigg(\frac{H_{0}^2 \Upsilon^{2
q+2}}{m^2}\bigg)^{-\frac{3 m}{8}} \bigg(6^{u+\frac{1}{2}} m^u (2
u+1) \sqrt{H_{0}^2 \Upsilon^{2 q+2}}
\\\nonumber
&\times \big(6^{\frac{3 m}{8}+2} \bigg\{-2 b-B_{2}-B_{1}
\sqrt{\rho_{0}}\bigg\}H_{0}^4 (3 m+4) \Upsilon^{4 q+4} \bigg(\frac{H_{0}^2
\Upsilon^{2 q+2}}{m^2}\bigg)^{\frac{3 m}{8}}-6^{\frac{3 m}{8}+\frac{1}{2}}
\\\nonumber
&\times(3 m+4) \sqrt{H_{0}^2 \Upsilon^{2 q+2}} \big(H_{0}
\Upsilon^{q+1} +6 H_{0}^2 \big(6 H_{0}^2 \Upsilon^{2 q+2}+2 H_{0}
\Upsilon^{2 q+2}\big) \Upsilon^{2 q+2}
\\\nonumber
&-36 H_{0}^4 \Upsilon^{4 q+4}\big)\bigg(\frac{H_{0}^2 \Upsilon^{2
q+2}}{m^2}\bigg)^{\frac{3 m}{8}} +6^{\frac{3 m}{8}} a
\sqrt{\rho_{0}} \big(3 H_{0} m (3 m+8) \Upsilon^{q+1}-48 H_{0}^2
\\\nonumber
&\times\big(12 H_{0}^2 \Upsilon^{2 q+2}-3 H_{0} m \Upsilon^{2
q+2}\big) \Upsilon^{2 q+2}-432 H_{0}^4 m \Upsilon^{4
q+4}\big)\big)-2^{\frac{3 m}{8}+u+\frac{7}{2}} 3^{\frac{3
m}{8}+u+\frac{3}{2}}
\\\nonumber
&\times(3 m+4) \big(H_{0}^2 \Upsilon^{2 q+2}\big)^{u+\frac{3}{2}}
\bigg(\frac{H_{0}^2 \Upsilon^{2 q+2}}{m^2}\bigg)^{\frac{3 m}{8}}
\big(-4 H_{0} u (u+1) \Upsilon^{q+1}+36 H_{0}^4 \Upsilon^{4 q+4}
\\\nonumber
&-72 H_{0}^4 (u+1) \Upsilon^{4 q+4}+24 h^3 (u+1) \Upsilon^{4
q+4}\bigg) g_1\bigg)\bigg[2^{-\frac{3 m}{8}-u-\frac{7}{2}}
3^{-\frac{3 m}{8}-u-\frac{5}{2}} m^{-u}
\\\nonumber
&\times\big(H_{0}^2 \Upsilon^{2 \Upsilon^{2 q+2}}{m^2}\big)^{-\frac{3 m}{8}}
\big(-2^{\frac{3 m}{8}+u+\frac{7}{2}} 3^{\frac{3 m}{8}+u+\frac{3}{2}} (3 m+4)
\big(H_{0}^2 \Upsilon^{2 q+2}\big)^{u+\frac{3}{2}} \big(12 H_{0}^2
\\\nonumber
&\times\Upsilon^{2 q+2}-12 H_{0}^2 (u+1) \Upsilon^{2 q+2}+4 H_{0}
(u+1) \Upsilon^{2 q+2}\big)\bigg(\frac{H_{0}^2 \Upsilon^{2
q+2}}{m^2}\bigg)^{\frac{3 m}{8}}
\\\nonumber
& -2^{\frac{3 m}{8}+u+\frac{5}{2}} 3^{\frac{3 m}{8}+u+\frac{1}{2}}
(3 m+4) \bigg(u+\frac{3}{2}\bigg) \big(H_{0}^2 \Upsilon^{2
q+2}\big)^{u+\frac{1}{2}} \big(-4 H_{0} u (u+1) \Upsilon^{q+1}
\\\nonumber
&\times \Upsilon^{4 q+4}-72 H_{0}^4 (u+1) \Upsilon^{4 q+4}+24
H_{0}^3 (u+1) \Upsilon^{4 q+4}\big) g_1 \bigg(\frac{H_{0}^2
\Upsilon^{2 q+2}}{m^2}\bigg)^{\frac{3 m}{8}}
\\\nonumber
& -\bigg[2^{\frac{3 m}{8} +u-\frac{1}{2}} 3^{\frac{3
m}{8}+u+\frac{3}{2}} (3 m+4) \big(H_{0}^2 \Upsilon^{2
q+2}\big)^{u+\frac{3}{2}}\bigg(\frac{H_{0}^2 \Upsilon^{2
q+2}}{m^2}\bigg)^{\frac{3 m}{8}-1} \bigg(-4 H_{0} u
\\\nonumber
&+36 H_{0}^4 \Upsilon^{4 q+4} -72 H_{0}^4 (u+1) \Upsilon^{4 q+4}+24
H_{0}^3 (u+1) \Upsilon^{4 q+4}\big) g_1\bigg]\bigg[m\bigg]
\\\nonumber
&+\bigg[2^{u-\frac{3}{2}} 3^{u-\frac{1}{2}} m^u (2 u+1) \big(6^{\frac{3 m}{8}+2}
\big(-2 b-B_{2}-B_{1} \sqrt{-H_{0} \Upsilon^{2 q+2}}+ H_{0}^4 \Upsilon^{4 q+4}
\\\nonumber
&\times(3 m+4) \bigg(\frac{H_{0}^2 \Upsilon^{2
q+2}}{m^2}\bigg)^{\frac{3 m}{8}}-6^{\frac{3 m}{8}+\frac{1}{2}} (3
m+4) \sqrt{H_{0}^2 \Upsilon^{2 q+2}} \bigg(H_{0} \Upsilon^{q+1}
\\\nonumber
&+6 H_{0}^2 \big(6 H_{0}^2 \Upsilon^{2 q+2}+2 H_{0} \Upsilon^{2 q+2}\big) \Upsilon^{2
q+2}-36 h^4 \Upsilon^{4 q+4}\big) g_2 \bigg(\frac{H_{0}^2 \Upsilon^{2
q+2}}{m^2}\bigg)^{\frac{3 m}{8}}
\\\nonumber
&+6^{\frac{3 m}{8}} \sqrt{\rho_{0}} \bigg(3H_{0} m (3 m+8) \Upsilon^{q+1} -48 H_{0}^2
\bigg(12 H_{0}^2 \Upsilon^{2 q+2}-3 H_{0} m \Upsilon^{2 q+2}\bigg) \Upsilon^{2 q+2}
\\\nonumber
&-432 H_{0}^4 m \Upsilon^{4 q+4}\bigg)\bigg)\bigg]\bigg[\sqrt{H_{0}^2 \Upsilon^{2
q+2}}\bigg]^{-1} +6^{u+\frac{1}{2}} m^u (2 u+1) \sqrt{H_{0}^2 \Upsilon^{2 q+2}}
\big(2^{\frac{3 m}{8}+2}
\\\nonumber
&\times3^{\frac{3 m}{8}+1} \big(-2 b-B_{2}-B_{1} \sqrt{-H_{0} \Upsilon^{2 q+2}}\bigg)
H_{0}^2 (3 m+4) \Upsilon^{2 q+2} \bigg(\frac{H_{0}^2 \Upsilon^{2
q+2}}{m^2}\bigg)^{\frac{3 m}{8}}
\\\nonumber
&-6^{\frac{3 m}{8}+\frac{1}{2}} (3 m+4) \sqrt{H_{0}^2 \Upsilon^{2
q+2}} \bigg(6 H_{0}^2 \Upsilon^{2 q+2}+2 H_{0} \Upsilon^{2
q+2}\bigg) g_2 \bigg(\frac{H_{0}^2 \Upsilon^{2
q+2}}{m^2}\bigg)^{\frac{3 m}{8}}
\\\nonumber
&+2^{\frac{3 m}{8}-2} 3^{\frac{3 m}{8}+2} \big(-2 b-B_{2}-B_{1} \sqrt{-H_{0}
\Upsilon^{2 q+2}}H_{0}^4 (3 m+4) \Upsilon^{4 q+4} \big(H_{0}^2 \Upsilon^{2 q+2}
\\\nonumber
&-72 H_{0}^2 m \Upsilon^{2 q+2}-8 \big(12 H_{0}^2 \Upsilon^{2 q+2}-3 h m \Upsilon^{2
q+2}\big)\big) -\bigg[2^{\frac{3 m}{8}-\frac{3}{2}} 3^{\frac{3 m}{8} -\frac{1}{2}}
\\\nonumber
&\times (3 m+4) \bigg(\frac{H_{0}^2 \Upsilon^{2 q+2}}{m^2}\bigg)^{\frac{3 m}{8}}
\big(H_{0} \Upsilon^{q+1}+6 H_{0}^2 \big(6 H_{0}^2 \Upsilon^{2 q+2}+2 H_{0}
\Upsilon^{2 q+2}\big) \Upsilon^{2 q+2}
\\\nonumber
&-36 H_{0}^4 \Upsilon^{4 q+4}\big) g_2\bigg]\bigg[\sqrt{H_{0}^2 \Upsilon^{2
q+2}}\bigg]^{-1} -\bigg[2^{\frac{3 m}{8} -\frac{7}{2}} 3^{\frac{3 m}{8} +\frac{1}{2}}
(3 m+4) \sqrt{H_{0}^2 \Upsilon^{2 q+2}}
\\\nonumber
&\times \bigg(\frac{H_{0}^2 \Upsilon^{2 q+2}}{m^2}\bigg)^{\frac{3 m}{8}-1} \big(H_{0}
\Upsilon^{q+1} +6 H_{0}^2 \big(6 H_{0}^2 \Upsilon^{2 q+2} +2 H_{0} \Upsilon^{2
q+2}\big) \Upsilon^{2 q+2} \bigg]\bigg[m\bigg]^{-1}\bigg]
\\\nonumber
&\times\bigg[(3 m+4) (2 u+1) \bigg(H_{0}^2 \Upsilon^{2 q+2}\bigg)^{5/2}
\bigg({\rho_{0}}+\frac{1}{2} \bigg(4  \sqrt{\rho_{0}} \bigg(\frac{H_{0}^2 \Upsilon^{2
q+2}}{m^2}\bigg)^{-\frac{3 m}{8}}
\\\nonumber
&+3 b+B_{2}+2 b+\sqrt{\rho_{0}} (2 \rho_{0}+1)-24 m^{-u} \big(H_{0}^2 \Upsilon^{2
q+2}\big)^{u+1} g_1+B_{1}\sqrt{-H_{0} \Upsilon^{2 q+2}}\bigg]^{-1}
\\\nonumber
& -\bigg[2^{-\frac{3 m}{8}-u-\frac{15}{2}} 3^{-\frac{3 m}{8}-u-\frac{5}{2}} m^{-u-1}
\bigg(\frac{H_{0}^2 \Upsilon^{2 q+2}}{m^2}\bigg)^{-\frac{3 m}{8}-1}
\big(6^{u+\frac{1}{2}} m^u (2 u+1) \sqrt{H_{0}^2 \Upsilon^{2 q+2}}
\\\nonumber
&\times \big(6^{\frac{3 m}{8}+2} \big(-2 b-B_{2}-B_{1} \sqrt{-H_{0} \Upsilon^{2
q+2}}\big) H_{0}^4 (3 m+4) \Upsilon^{4 q+4} \bigg(\frac{h^2 \Upsilon^{2
q+2}}{m^2}\bigg)^{\frac{3 m}{8}}
\\\nonumber
&-6^{\frac{3m}{8}+\frac{1}{2}} (3 m+4) \sqrt{H_{0}^2 \Upsilon^{2 q+2}} \big(H_{0}
\Upsilon^{q+1}+6 H_{0}^2 \big(6 H_{0}^2 \Upsilon^{2 q+2}+2 H_{0} \Upsilon^{2
q+2}\big) \Upsilon^{2 q+2}
\\\nonumber
&-36 H_{0}^4\Upsilon^{4 q+4}\big) g_2 \bigg(\frac{H_{0}^2 \Upsilon^{2
q+2}}{m^2}\bigg)^{\frac{3 m}{8}} +6^{\frac{3 m}{8}} a \sqrt{\rho_{0}} \bigg(3 H_{0} m
(3 m+8) \Upsilon^{q+1}-48 h^2
\\\nonumber
&\times \big(12 H_{0}^2 \Upsilon^{2 q+2}-3 H_{0} m \Upsilon^{2 q+2}\big) \Upsilon^{2
q+2} -432 H_{0}^4 m \Upsilon^{4 q+4}\big)\big) -2^{\frac{3 m}{8} +u+\frac{7}{2}}
3^{\frac{3 m}{8} +u+\frac{3}{2}}
\\\nonumber
&\times(3 m+4) \big(H_{0}^2 \Upsilon^{2 q+2}\big)^{u+\frac{3}{2}}
\bigg(\frac{H_{0}^2 \Upsilon^{2 q+2}}{m^2}\bigg)^{\frac{3 m}{8}}
\big(-4 H_{0} u (u+1) \Upsilon^{q+1} +36 H_{0}^4 \Upsilon^{4 q+4}
\\\nonumber
&-72 H_{0}^4 (u+1) \Upsilon^{4 q+4}+24 H_{0}^3 (u+1) \Upsilon^{4 q+4}\bigg)
g_1\bigg)\bigg]\bigg[(3 m+4) (2 u+1) \big(H_{0}^2 \Upsilon^{2 q+2}\big)^{5/2}
\\\nonumber
&\times \bigg(\rho_{0}+\frac{1}{2} \bigg(4\sqrt{\rho_{0}} \bigg(\frac{H_{0}^2
\Upsilon^{2 q+2}}{m^2}\bigg)^{-\frac{3 m}{8}}+3 b+B_{2}+2 b+\sqrt{\rho_{0}} (2 +1)-24
m^{-u}
\\\nonumber
&\times \big(H_{0}^2 \Upsilon^{2 q+2}\big)^{u+1} g_1+B_{1} \sqrt{-H_{0} \Upsilon^{2
q+2}}\bigg)\bigg)\bigg]^{-1}-\bigg[5\ 2^{-\frac{3 m}{8}-u-\frac{11}{2}} 3^{-\frac{3
m}{8}-u-\frac{7}{2}}
\\\nonumber
&\times m^{-u}\bigg(\frac{H_{0}^2 \Upsilon^{2
q+2}}{m^2}\bigg)^{-\frac{3 m}{8}} \big(6^{u+\frac{1}{2}} m^u (2 u+1)
\sqrt{H_{0}^2 \Upsilon^{2 q+2}} \big(6^{\frac{3 m}{8}+2} \bigg(-2
b-\text{B2}
\\\nonumber
& -B_{1} \sqrt{-H_{0} \Upsilon^{2 q+2}}\bigg) H_{0}^4 (3 m+4) \Upsilon^{4 q+4}
\bigg(\frac{H_{0}^2 \Upsilon^{2 q+2}}{m^2}\bigg)^{\frac{3 m}{8}}-6^{\frac{3
m}{8}+\frac{1}{2}} (3 m+4)
\\\nonumber
&\times \sqrt{H_{0}^2 \Upsilon^{2 q+2}} \big(H_{0} \Upsilon^{q+1} +6 H_{0}^2 \big(6
H_{0}^2 \Upsilon^{2 q+2}+2 h \Upsilon^{2 q+2}\big) \Upsilon^{2 q+2}-36 H_{0}^4
\Upsilon^{4 q+4}\big) g_2
\\\nonumber
&\times \bigg(\frac{H_{0}^2 \Upsilon^{2 q+2}}{m^2}\bigg)^{\frac{3 m}{8}} +6^{\frac{3
m}{8}}  \sqrt{\rho_{0}} \big(3 H_{0} m (3 m+8) \Upsilon^{q+1}-48 H_{0} ^2 \big(12
H_{0}^2 \Upsilon^{2 q+2}
\\\nonumber
&-3 H_{0} m \Upsilon^{2 q+2}\big) \Upsilon^{2 q+2} -432 H_{0}^4 m \Upsilon^{4
q+4}\big)\big) \bigg(\frac{H_{0}^2 \Upsilon^{2 q+2}}{m^2}\bigg)^{\frac{3 m}{8}}
-2^{\frac{3 m}{8}+u+\frac{7}{2}} 3^{\frac{3 m}{8}+u+\frac{3}{2}}
\\\nonumber
&\times (3 m+4) \big(H_{0}^2 \Upsilon^{2 q+2}\big)^{u+\frac{3}{2}}
\bigg(\frac{H_{0}^2 \Upsilon^{2 q+2}}{m^2}\bigg)^{\frac{3 m}{8}}
\big(-4 H_{0} u (u+1) \Upsilon^{q+1} +36 H_{0}^4 \Upsilon^{4 q+4}
\\\nonumber
&-72 H_{0}^4 (u+1) \Upsilon^{4 q+4}+24 H_{0}^3 (u+1) \Upsilon^{4 q+4}\bigg)
g_1\bigg)\bigg]\bigg[(3 m+4) (2 u+1) \big(H_{0}^2 \Upsilon^{2 q+2}\big)^{7/2}
\\\nonumber
&+\frac{1}{2} \bigg(4  \sqrt{\rho_{0}} \bigg(\frac{H_{0}^2 \Upsilon^{2
q+2}}{m^2}\bigg)^{-\frac{3 m}{8}}+3 b+B_{2}+2 b+\sqrt{\rho_{0}} (2\rho_{0}+1)-
\big(H_{0}^2 \Upsilon^{2 q+2}\big)^{u+1} g_1
\\\nonumber
&\times \sqrt{-H_{0} \Upsilon^{2 q+2}}\bigg]^{-1}-\bigg[2^{-\frac{3
m}{8}-u-\frac{9}{2}} 3^{-\frac{3 m}{8}-u-\frac{5}{2}} m^{-u}
-\bigg\{\bigg(\frac{H_{0}^2 \Upsilon^{2 q+2}}{m^2}\bigg)^{-\frac{3 m}{8}-1}\bigg\}
\\\nonumber
&\times\big\{4 m\big\} -4 m^{-u} (u+1) \big(H_{0}^2 \Upsilon^{2
q+2}\bigg)^u g_1\big) \big(6^{u+\frac{1}{2}} m^u (2 u+1) \sqrt{h^2
\Upsilon^{2 q+2}} \big(6^{\frac{3 m}{8}+2}
\\\nonumber
&-B_{1} \sqrt{-H_{0} \Upsilon^{2 q+2}}\bigg) H_{0}^4 (3 m+4) \Upsilon^{4 q+4}
\bigg(\frac{H_{0}^2 \Upsilon^{2 q+2}}{m^2}\bigg)^{\frac{3 m}{8}}-6^{\frac{3
m}{8}+\frac{1}{2}} (3 m+4)
\\\nonumber
&\times \sqrt{H_{0}^2 \Upsilon^{2 q+2}} \big(H_{0} \Upsilon^{q+1} +6 H_{0}^2 \big(6
H_{0}^2 \Upsilon^{2 q+2}+2 H_{0} \Upsilon^{2 q+2}\big) \Upsilon^{2 q+2}-36 H_{0}^4
\Upsilon^{4 q+4}\big) g_2
\\\nonumber
&\times \bigg(\frac{H_{0}^2 \Upsilon^{2 q+2}}{m^2}\bigg)^{\frac{3 m}{8}} +6^{\frac{3
m}{8}}  \sqrt{\rho_{0}} \big(3 H_{0} m (3 m+8) \Upsilon^{q+1}-48 H_{0}^2 \big(12
H_{0}^2 \Upsilon^{2 q+2}
\\\nonumber
&-3 H_{0} m \Upsilon^{2 q+2}\big) \Upsilon^{2 q+2} -H_{0}^4 m \Upsilon^{4
q+4}\big)\big)-2^{\frac{3 m}{8}+u+\frac{7}{2}} 3^{\frac{3 m}{8}+u+\frac{3}{2}} (3
m+4) \big(H_{0}^2 \Upsilon^{2 q+2}\big)^{u+\frac{3}{2}}
\\\label{66}
&\times \bigg(\frac{H_{0}^2 \Upsilon^{2 q+2}}{m^2}\bigg)^{\frac{3
m}{8}}\big(-4 H_{0} u (u+1) \Upsilon^{q+1}+36 H_{0}^4 \Upsilon^{4
q+4}-72 H_{0}^4 (u+1) \Upsilon^{4 q+4}\bigg].
\end{align}
\\
\textbf{Data Availability Statement:} No new data were generated or
analyzed in support of this research.

\end{document}